\documentclass[conference,letterpaper,10pt]{IEEEtran}

\usepackage{framed}
\usepackage{tcolorbox}
\usepackage{amsmath}
\usepackage{mathtools}
\usepackage{bbm}
\usepackage[ruled,linesnumbered]{algorithm2e}
\usepackage{bm}
\usepackage{stackrel}
\usepackage{subfig}
\usepackage{epsf}
\usepackage{amsmath,amssymb}

\usepackage{graphicx}
\usepackage{color}
\usepackage{cite}
\usepackage{multirow,tabularx}
\usepackage{ifthen}
\usepackage{epstopdf}
\input{epstopdf.sty}

\usepackage{hyperref}
\hypersetup{
	colorlinks=true, 
    linkcolor=blue,  
    citecolor=blue,  
    filecolor=blue,  
    urlcolor=blue    
}
\usepackage{times}
\input{epsf.sty}

\makeatletter
\def\BState{\State\hskip-\ALG@thistlm}
\makeatother

\newcommand{\hide}[1]{\ifthenelse{\boolean{false}}{#1}{}}

\newtheorem{theorem}{{\bf Theorem}}

\newtheorem{lemma}{{\bf Lemma}}

\newcommand{\qed}{\nobreak \ifvmode \relax \else
      \ifdim\lastskip<1.5em \hskip-\lastskip
      \hskip1.5em plus0em minus0.5em \fi \nobreak
      \vrule height0.75em width0.5em depth0.25em\fi}

\newcommand{\beq}{\begin{equation}}
\newcommand{\eeq}{\end{equation}}
\newcommand{\barr}{\begin{array}}
\newcommand{\earr}{\end{array}}

\newcommand{\benum}{\begin{enumerate}}
\newcommand{\eenum}{\end{enumerate}}

\newcommand{\bit}{\begin{itemize}}
\newcommand{\eit}{\end{itemize}}

\newcommand{\bc}{\begin{center}}
\newcommand{\ec}{\end{center}}

\newcommand{\bdes}{\begin{description}}
\newcommand{\edes}{\end{description}}

\newcommand{\bfig}{\begin{figure}}
\newcommand{\efig}{\end{figure}}

\newcommand{\bemq}{\begin{quote} \begin{em}}
\newcommand{\eemq}{\end{em} \end{quote}}

\newcommand{\bmp}{\begin{minipage}}
\newcommand{\emp}{\end{minipage}}

\newcommand{\bsp}{\begin{slide*}}
\newcommand{\esp}{\end{slide*}}
\newcommand{\bsl}{\begin{slide}}
\newcommand{\esl}{\end{slide}}

\newcommand{\blem}{\begin{lemma}}
\newcommand{\elem}{\end{lemma}}
\newcommand{\bthm}{\begin{theorem}}
\newcommand{\ethm}{\end{theorem}}

\IEEEoverridecommandlockouts

\begin{document}

\title{
WiSwarm: Age-of-Information-based Wireless Networking for Collaborative Teams of UAVs\color{black}}
\author{Vishrant Tripathi*, Igor Kadota*, Ezra Tal*, Muhammad Shahir Rahman, Alexander Warren,\\Sertac Karaman, and Eytan Modiano
\IEEEcompsocitemizethanks{\IEEEcompsocthanksitem Vishrant Tripathi, Ezra Tal, Muhammad Shahir Rahman, Alexander Warren, Sertac Karaman, and Eytan Modiano are with the Laboratory for Information and Decision Systems (LIDS), Massachusetts Institute of Technology, Cambridge, MA, 02139. Igor Kadota is with the Department of Electrical Engineering, Columbia University, New York, NY, 10027. This work was supported in part by the NSF under grant CNS-1713725 and the ARO under grant W911NF1910322. A version of this work will be presented at IEEE INFOCOM 2023.\protect\\
E-mail: \{vishrant, eatal, shahir, warrena, sertac, modiano\}@mit.edu and igor.kadota@columbia.edu.\protect\\
*These authors contributed equally to this work
}
}

\IEEEaftertitletext{\vspace{-0.6\baselineskip}}

\maketitle
\begin{abstract}
%
The Age-of-Information (AoI) metric has been widely studied in the theoretical communication networks and queuing systems literature. However, experimental evaluation of its applicability to complex real-world time-sensitive systems is largely lacking. In this work, we develop, implement, and evaluate an AoI-based application layer middleware that enables the customization of WiFi networks to the needs of time-sensitive applications. By controlling the storage and flow of information in the underlying WiFi network, our middleware can: (i) prevent packet collisions; (ii) discard stale packets that are no longer useful; and (iii) dynamically prioritize the transmission of the most relevant information. To demonstrate the benefits of our middleware, we implement a mobility tracking application using a swarm of UAVs communicating with a central controller via WiFi. Our experimental results show that, when compared to WiFi-UDP/WiFi-TCP, the middleware can improve information freshness by a factor of 109x/48x and tracking accuracy by a factor of 4x/6x, respectively. Most importantly, our results also show that the performance gains of our approach increase as the system scales and/or the traffic load increases.  
\end{abstract}

\section{Introduction}\label{sec.Intro}

Emerging time-sensitive applications increasingly rely on collaborative multi-agent systems. Examples are abundant: search and rescue missions using a team of unmanned aerial vehicles (UAVs), smart factories with connected automated machinery, and smart city intersections with connected self-driving cars. In such application domains, it is essential that agents communicate in a timely manner about changes in the environment and adapt their behavior accordingly. A major roadblock in deploying these applications in the real-world is that traditional communication networks were not designed to support large-scale multi-agent system that need to share time-sensitive information to collaborate effectively.

WiFi is a common choice for deploying time-sensitive applications. Some examples include: automated fulfilment warehouses at Amazon \cite{KivaWiFi}, vehicle-to-everything communication in New York City \cite{VehicleToInfrastructureDSRC,AppLevelDSRC,NYCDOT}, and various multi-agent systems using teams of UAVs and ground robots \cite{CERBERUS,SAR,DistributedRobotFormation,CooperativeSLAM,ASTRO,DroneCinema,DOOR-SLAM,BeeCluster,MultiRobotSlam,MultiRobotMapping,hu2020hivemind}.  
WiFi is attractive for deploying such systems because it is inexpensive, tried-and-true, and readily available in sensors, cameras, UAVs, and robotic platforms. However, it is well-known that WiFi's performance degrades sharply as the network size scales and traffic load increases. This is due to WiFi's Carrier-Sense Multiple Access (CSMA) distributed random access mechanism that works well for small-scale underloaded networks, but not for large-scale systems with stringent latency or freshness requirements. When a larger number of sources attempt to transmit using distributed random access, the higher probability of packet collisions leads to lower throughput and higher latency, which can result in degraded performance (or even failure) of the time-sensitive application.

\textbf{Our contributions: (1) Middleware design.} We develop a networking middleware that makes WiFi networks customizable, allowing system designers to easily tailor WiFi to the needs of specific time-sensitive applications. 
Our middleware drives the underlying distributed WiFi network to behave as a network with centralized resource allocation and with custom queues at the sources. By controlling the storage and flow of information in the WiFi network, the middleware: (i) prevents packet collisions; (ii) dynamically prioritizes the transmissions that are most valuable to the application; and (iii) discards stale packets that are no longer useful to the application before they are ever transmitted, thus alleviating congestion.

The networking middleware has two distinct features. First, it is \emph{implemented at the application layer}, without (any) modifications to lower layers of the networking protocol stack. The middleware runs over UDP/IP and standard 802.11 WiFi, making it easy to customize and integrate to existing time-sensitive applications that are already implemented using WiFi, such as \cite{KivaWiFi,VehicleToInfrastructureDSRC,AppLevelDSRC,NYCDOT,CERBERUS,SAR,DistributedRobotFormation,CooperativeSLAM,ASTRO,DroneCinema,BeeCluster,DOOR-SLAM,MultiRobotSlam,MultiRobotMapping,hu2020hivemind}. Second, the middleware is \emph{designed around the idea of information freshness}, specifically the \textbf{Age-of-Information (AoI) metric}. The AoI captures the freshness of the information \emph{from the perspective of the destination}, in contrast to the long-established packet delay that represents the latency of a \emph{particular packet}. The networking middleware can leverage AoI to prioritize transmissions to destinations with stale information. 
Keeping information fresh is critical for time-sensitive applications, especially those that rely on cooperative multi-agent systems. 

\textbf{Our contribution: (2) WiSwarm implementation.} To demonstrate the performance improvement that can be achie\-ved by customizing the WiFi network, we implement Wi\-Swarm: an instantiation of our networking middleware for a mobility tracking application that relies on a collaborative UAV swarm. Following the recent growing interest in computational offloading to enhance the scale of multi-agent robotics applications \cite{chinchali2021network}, we implement a mobility tracking 
application composed of several small and inexpensive UAVs and one leader node with high compute power. Each UAV senses the environment (e.g., collects video) and transmits this contextual information to the leader node. The leader node consolidates the information from the UAVs and transmits trajectory updates that allow the UAVs to track the moving objects. Clearly, \emph{it is essential to keep the contextual information at the leader node and the trajectory updates at every UAV as fresh as possible}, since outdated information loses its value and can lead to system failures (e.g., UAV losing track of an object) and safety risks (e.g., UAV collisions). 

\begin{figure}
	\centering
	\includegraphics[width=\linewidth]{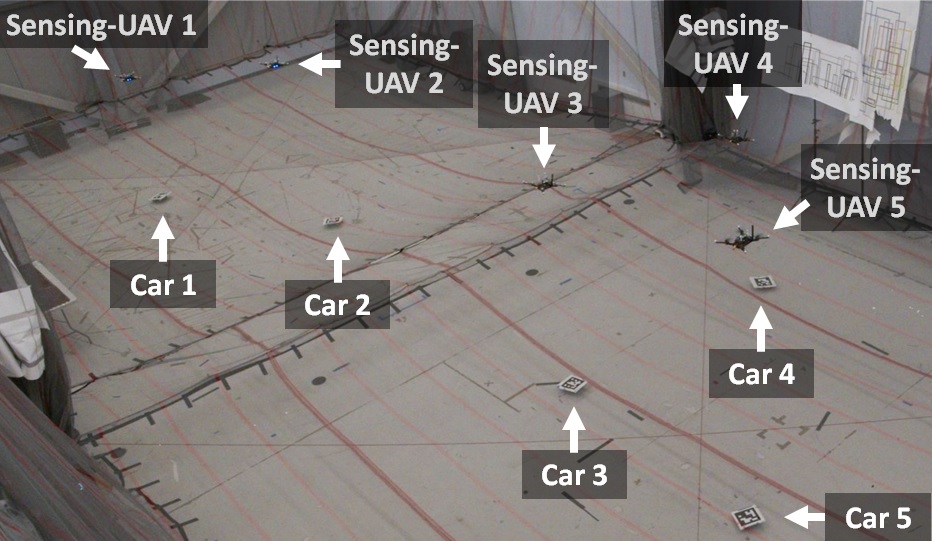}
	\caption{Flight experiment with five UAVs.}
	\label{fig.flight_setup}
\end{figure}
We evaluate WiSwarm in \textbf{flight experiments}
with up to five sensing-UAVs (see Fig.~\ref{fig.flight_setup}), and in \textbf{stationary experiments} with up to fourteen  Raspberry Pis (RasPis) emulating UAVs. We collect data from nearly 4 hours of flight tests and $400$ hours of stationary tests. We also provide a video \cite{Video} summarizing the setup and results from our flight experiments. Our experimental results show that WiSwarm significantly outperforms WiFi in terms of throughput, information freshness, tracking performance, and scalability. 
The stationary experiments with fourteen sources 
shows that WiSwarm improves information freshness by a factor of $109$, and tracking error by a factor of $4$. The flight tests show that mobility tracking with WiFi can support \emph{at most two} sensing-UAVs while WiSwarm can support \emph{at least five} sensing-UAVs under similar conditions. 

\textbf{To the best of our knowledge, this is the first work to develop and implement an application-layer solution that optimizes information freshness in the wireless network without requiring modifications to lower layers of the networking protocol stack and, also, the first work to experimentally evaluate the impact of an AoI-based solution in a real-world time-sensitive application.}

The remainder of this work is organized as follows. 
In Sec.~\ref{sec.Middleware}, we introduce the AoI metric and describe the middleware. 
In Sec.~\ref{sec.WiSwarm}, we describe the design and implementation of WiSwarm. 
In Sec.~\ref{sec.Evaluation}, we evaluate the performance of WiSwarm in flight tests and stationary experiments. 
Finally, in Sec.~\ref{sec.Conclusion}, we conclude and discuss future work.

\section{Networking Middleware for Information Freshness}\label{sec.Middleware}


\begin{figure}
	\centering
	\includegraphics[width=\linewidth]{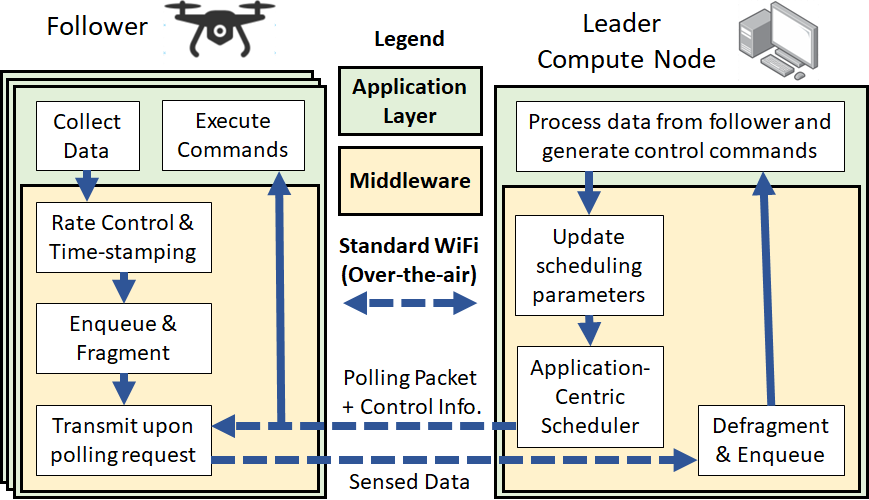}
	\caption{AoI-based application layer Networking Middleware.}
	\label{fig.overview}
\end{figure}


In this section, we describe a networking middleware (illustrated in Fig.~\ref{fig.overview}) that customizes WiFi to the needs of the important and broad class of time-sensitive applications that rely on multi-agent systems. In these applications, agents (also called followers) collect and transmit time-sensitive information to a central compute node (also called the leader). 
The leader consolidates the received information and coordinates the followers' behavior in a timely manner. Naturally, it is critical to keep information in the network as fresh as possible. 
%
We formally define the Age-of-Information metric in Sec.~\ref{sec.AoI}. In Sec.~\ref{sec.MiddlewareDesign}, we describe the middleware design based on the considerations in Secs.~\ref{sec.AoI}, \ref{sec.Queueing}, and~\ref{sec.Scheduler}.

\subsection{Age-of-Information Metric}
\label{sec.AoI}
\begin{figure}
	\centering
	\includegraphics[width=0.95\linewidth]{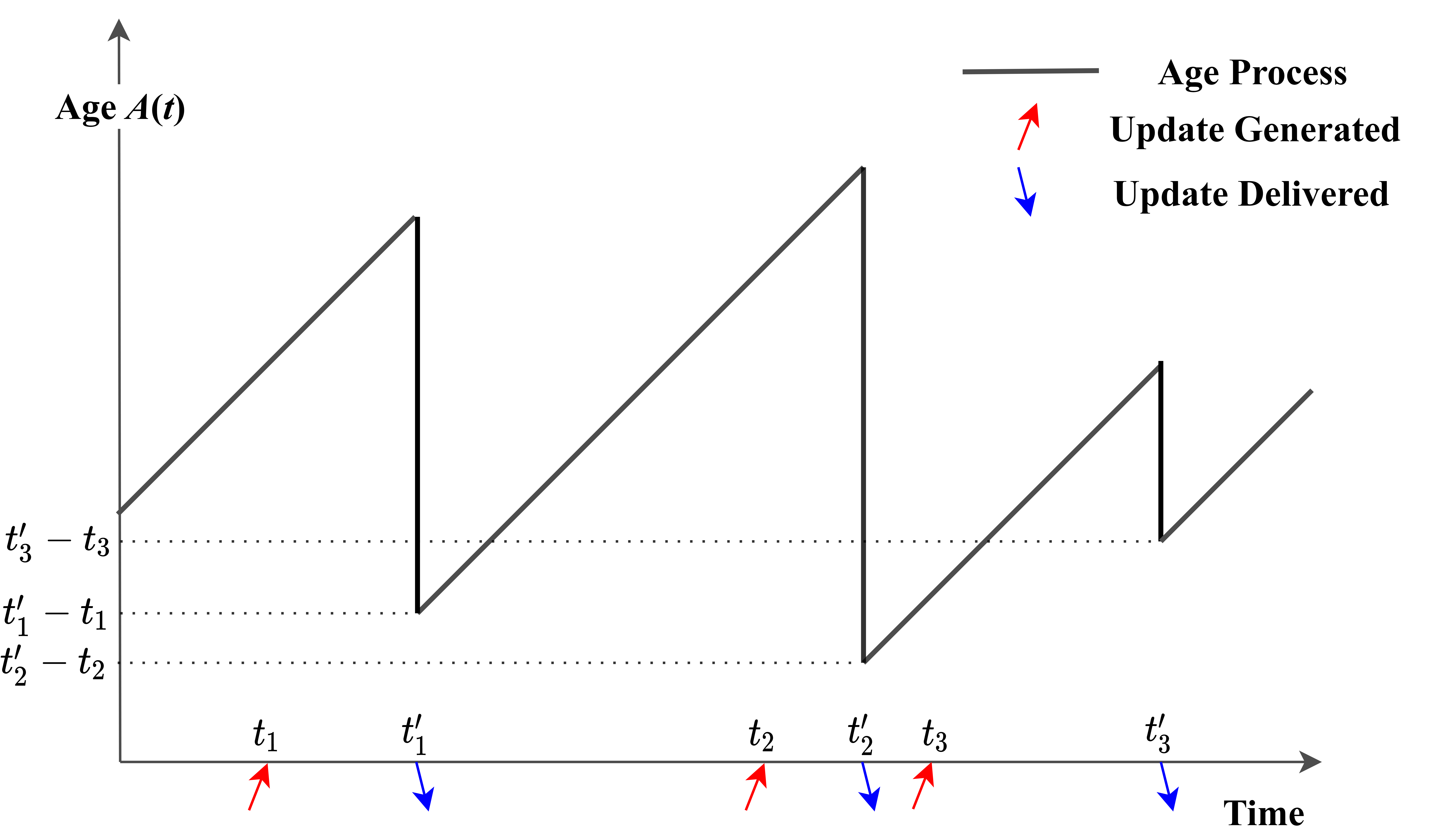}
	\caption{AoI evolution.} 
	\label{fig.AoI}
\end{figure}

AoI is an end-to-end metric that characterizes \emph{how old the information is from the perspective of the destination} \cite{kaul2012real}. Consider a multi-agent system in which updates from followers are time-stamped upon generation. Let $\tau_i(t)$ be the \emph{largest time-stamp} of an update from follower~$i$ received by the leader by time~$t$. The AoI associated with follower~$i$ is defined as $A_i(t):=t-\tau_i(t)$. The AoI increases linearly with time when no updates are delivered, representing the information getting older. At the moment a \emph{fresher} update from follower~$i$ is received by the leader, the value of $\tau_i(t)$ increases and the AoI reduces to the delay of the received update. This evolution of the AoI metric with time is illustrated in Fig.~\ref{fig.AoI}.

Over the past decade there has been a rapidly growing body of works analyzing AoI in different contexts (see surveys in \cite{yates2021age,kosta2017age,sun2019age_book}). Several theory-oriented papers have analyzed AoI in queuing systems \cite{kaul2012real, kam2013age, huang2015optimizing, inoue2018general, yin17_tit_update_or_wait, bedewy2019minimizing} and proposed novel network control mechanisms \cite{kadota2018scheduling2, talak2018optimizing, maatouk2020optimality, tripathi2019whittle, tripathi2021online,jhun2018age,farazi2018age} that could potentially be leveraged in real-world applications. 

More recently, a few works \cite{AoI_measure_1,AoI_measure_3,shreedhar2018acp,AoI_Wierman,AoI_SDR,ayan2021experimental,kadota2021age,kadota2021wifresh} have considered system implementations. These system-oriented works can be separated into two categories: (i) measurement of AoI in real networks 
\cite{AoI_measure_1,AoI_measure_3}; and (ii) evaluation of communication networks that attempt to minimize AoI by looking at congestion control \cite{shreedhar2018acp}, traffic engineering \cite{AoI_Wierman}, and medium access using Software Defined Radios (SDRs) \cite{ayan2021experimental,kadota2021age,kadota2021wifresh,AoI_SDR}. However, there has been no prior work on the experimental evaluation of the impact of an AoI-based networking solution in a real-world time-sensitive application, which is the focus of this work.

\subsection{Customizable Queueing at the Followers}\label{sec.Queueing}
Data generation and queueing have significant impact on information freshness. The \emph{follower middleware} architecture illustrated on the left in Fig.~\ref{fig.overview} receives updates at rates that are determined by sensors/applications, then it time-stamps and enqueues these updates. Upon receiving a polling request from the leader, the follower middleware releases a \emph{single} update via UDP/IP to lower layers of the network protocol stack. Our middleware incorporates two key ideas from the AoI literature to enable information freshness - a mechanism to control the update generation rate, and an implementation of Last-In First-Out (LIFO) queues.  

First-In First-Out (FIFO) queues are the default queuing implementation in most communication networks. 
However, to manage AoI, they require careful control of the arrival rate. If updates are generated at a low rate, then the information updates are too infrequent. On the other hand, if updates are generated at a very high rate, then the FIFO queue will often be backlogged and fresh updates will have to face large queueing delays. To address this problem, we implement a rate control mechanism at the followers that can be used when applications use FIFO queues.

\textbf{Rate Control}: To adjust the update generation rate, the rate control mechanism 
only updates its queue at fixed intervals of time, dropping any updates generated in between. This mechanism ensures that the middleware only accepts new updates at the desired rate. 
Note that finding the optimal generation rate for a given network setup is a nontrivial task, as the optimal rate depends on the network's topology, traffic load, link reliability, and Medium Access Control (MAC) mechanism. To illustrate the impact of the generation rate on information freshness, we plot in Fig.~\ref{fig.plot_wifi_1}(a)  the AoI of a standard WiFi system (that uses FIFO queues at the MAC layer) with different update generation rates, including the optimal rate which is obtained by grid search.

\textbf{LIFO Queues:} Last-In First-Out (LIFO) queues transmit the most recently generated update first, making them ideal for applications that rely on the knowledge of the \emph{current state} of the system, such as mobility tracking. When an update is generated, the LIFO queue simply replaces the old head-of-line update with the fresh update. 
A higher update generation rate at the followers can only lead to fresher updates at  the leader and, hence, a lower AoI. 
LIFO queues have been shown to be \emph{optimal} for minimizing AoI in a wide variety of network settings \cite{bedewy2019minimizing,AoI_management}. However, LIFO queues are rarely implemented at the transport, MAC, or physical layers in practice. Our middleware supports both FIFO and LIFO queues at the application layer, while also supporting rate control, providing the system designer with two important tools to manage AoI. 

\subsection{Customizable Transmission Scheduling at the Leader}\label{sec.Scheduler}

The multiple access mechanism controls the method by which followers and leader share information using the limited communication resources. WiFi employs a distributed random access mechanism that works well for small-scale underloaded networks. However, for large-scale congested networks, it leads to excess packet collisions that in turn lead to lower throughput and higher latency, and ultimately poor performance for real-time applications.


We design the \emph{leader middleware}, illustrated on the right in Fig.~\ref{fig.overview}, to: (i) prevent packet collisions; (ii) enable dynamic prioritization of the transmissions that are most valuable to the application; and (iii) facilitate integration with existing multi-agent systems that use WiFi. 
The middleware drives the underlying distributed WiFi network to behave like a centralized network with support for polling. Specifically, the leader middleware coordinates the flow of information in the network by sending polling packets to the followers selected for transmission. At every decision time $t$, the leader selects the next follower to poll based on an application-centric \emph{transmission scheduling policy} $\pi$, which can be a function of the current AoI of the followers $A_i(t)$, the reliability of the WiFi links $p_i(t)$, where $p_i(t)\in(0,1]$ represents the probability of a successful transmission from follower $i$ to the leader, and the application-defined priority weights $w_i(t)\geq 0$, which represent the relative importance of each follower's information to the overall application goal. For example, in a mobility tracking application, the estimated velocities of the moving objects can be assigned as application weights $w_i(t)$, since faster objects may require more updates than slower objects in order to achieve the same tracking performance. 


To capture application priorities and information freshness, we define the expected time-average of the weighted sum of AoIs across the entire network as
\begin{equation}\label{eq:AoI_opt}
\frac{1}{T} \mathbb{E} \left[ \sum_{i=1}^{N} \left(\int_{t=0}^{T} w_i(t)A_i(t) dt \right) \right] \; ,
\end{equation}
where $N$ is the number of followers, $T$ is the time-horizon, and the expectation is with respect to the randomness in the link's reliability $p_i(t)$ and the policy $\pi$. 
%
%

Many theoretical works \cite{kadota2018scheduling2, talak2018optimizing, maatouk2020optimality,tripathi2019whittle} have studied the structure of scheduling policies that attempt to minimize objective functions of the form \eqref{eq:AoI_opt}. A key take away from these works is that, given the knowledge of the application weights $w_i(t)$, link reliabilities $p_i(t)$, and information freshness $A_i(t)$ of every follower $i$, \emph{the Whittle's Index Policy is a near-optimal solution to the problem of minimizing} \eqref{eq:AoI_opt}. The \textbf{Whittle's Index Policy} selects, at every decision time $t$, the follower $i^*$ that satisfies
\begin{equation}
    \label{eq:AoI_whittle}
	i^{*} \in \textstyle\operatorname{argmax}_i \left\{ w_i(t) p_i(t) A^2_i(t) \right\} \; ,
\end{equation}
with ties being broken arbitrarily. 
Intuitively, the Whittle's Index Policy is polling the followers associated with high application weights, reliable WiFi links, and outdated information at the leader. 

Recent works have developed similar Whittle's Index Policies to address generalizations of \eqref{eq:AoI_opt}. Specifically, \cite{tripathi2019whittle} addressed the problem of minimizing general non-decreasing cost functions of AoI, $f_i(A_i(t))$, as opposed to simply minimizing $A_i(t)$, and \cite{tripathi2021online} considered network settings with time-varying, unknown and even adversarial application weights $w_i(t)$. This suggests that Whittle's Index Policies are remarkably robust and can be applied to a wide variety of applications. Moreover, the Whittle's Index Policy has low computational complexity: it only requires solving the maximization in \eqref{eq:AoI_whittle} and computing estimates of $w_i(t)$, $p_i(t)$, and $A_i(t)$. 

\subsection{Middleware Design}\label{sec.MiddlewareDesign}

We describe the networking middleware illustrated in Fig.~\ref{fig.overview}, which incorporates both the application-centric queueing at the followers and transmission scheduling at the leader.
%



\textbf{Followers} collect information updates about their immediate environment (e.g., video, pictures, laser scans, and temperature) and about their own platforms (e.g., position, attitude, velocity, and battery level). These updates are sent to the \emph{follower middleware} to be prepared for transmission.

The rate control mechanism decides whether each update is discarded or enqueued. The follower middleware time-stamps each update that is not discarded at the time of collection and enqueues them. These time-stamps are used to compute $A_i(t)=t-\tau_i(t)$ at the leader upon delivery. 
The queuing discipline, update rates, and queue buffer sizes can be controlled by the middleware to satisfy the requirements of the application. 


When the follower receives a polling packet, 
it releases a single information update from its queue. Assuming that the update does not exceed the maximum length of the UDP payload (or any threshold set by the system designer), the released update can be simply forwarded via UDP/IP to lower layers of the networking protocol stack. However, if the update is too large, then the middleware divides the update into \emph{fragments}. 
Fragments are stored in a separate FIFO queue and then transmitted one-by-one to the leader. Each fragment is transmitted via UDP/IP over standard WiFi. Since the maximum WiFi frame length can be smaller than the UDP payload size, it is possible that WiFi will require multiple successful over-the-air transmissions to deliver a single fragment to the leader. If WiFi fails to deliver a fragment, the middleware attempts to re-transmit the fragment using an error-control mechanism based on acknowledgements at the fragment level.

\textbf{The Leader's} responsibilities include coordinating both the flow of information in the WiFi network and the followers' behavior. To do so, the leader manages the generation and transmission of \emph{polling packets} and \emph{control information}. Since follower's updates are transmitted only upon reception of a polling packet, the leader has almost full control of the flow of information in the WiFi network, irrespective of the number of followers and the amount of data they generate. 


The leader uses the Whittle's Index Policy \eqref{eq:AoI_whittle} to decide the next follower to poll. After transmitting a polling packet, the leader waits for the reception of a fragment. If this waiting period exceeds a timeout interval (e.g., $300$\thinspace{milliseconds}), the attempt is assumed to have failed. Upon receiving a fragment or after a timeout, the leader prepares for the transmission of the next polling packet.

Prior to transmitting the next polling packet, the leader takes a series of steps that depend on whether the received fragment was the final fragment of an information update or not. If the received fragment from follower $i$ was not the final one, then the leader middleware simply updates $p_i(t)$. On the other hand, if the received fragment was the final, then the leader: 
(i) updates $p_i(t)$; (ii) combines fragments to obtain the original information update; (iii) extracts the associated time-stamp and updates $A_i(t)$; (iv) sends the information update to the application for processing; and (v) updates both $w_i(t)$ and the \emph{control information} based on the results of this processing.

To estimate $p_i(t)$, the leader computes $\hat{p}_i(t)=D_i(t)/W,$ where $D_i(t)$ is the number of polling packets which received a successful response from follower $i$ out of the last $W$ polling packets sent to it. To accurately compute $A_i(t)=t-\tau_i(t)$, where $t$ is the current time measured by the leader and $\tau_i(t)$ is the largest time-stamp received from follower $i$, the clock at follower $i$ should be synchronized with the leader's clock. The middleware performs periodic clock synchronization across all followers and the leader, at every $120$ seconds using NTP \cite{NTP}. 

After performing the necessary updates, the leader middleware transmits a new polling packet to the selected follower. The latest control information is broadcast to all followers along with every polling packet. 

\section{WiSwarm: Design and Implementation}\label{sec.WiSwarm}

In this section, we describe the design and implementation of WiSwarm which is an instantiation of the networking middleware for information freshness discussed in Sec.~\ref{sec.Middleware} tailored to a mobility tracking application. 
\subsection{Mobility Tracking Application}\label{sec.MobilityTracking}
Consider a setting where multiple UAVs are tracking moving objects on the ground. Clearly, outdated information about the position of the objects has a direct impact on the tracking capability of the UAVs. Ideally, the UAVs would like to receive fresh information about the objects continuously. One simple system design that achieves this goal consists of UAVs with high on-board computational power that are able to process video frames acquired from their cameras to detect and track objects. The continuous stream of images is processed locally, adding almost no delay, which keeps the UAVs updated about the position of the objects. A critical drawback of this approach is the prohibitively high cost of deploying numerous UAVs with high on-board computational power. 
\begin{figure}
	\centering
	\includegraphics[width=1.0\linewidth]{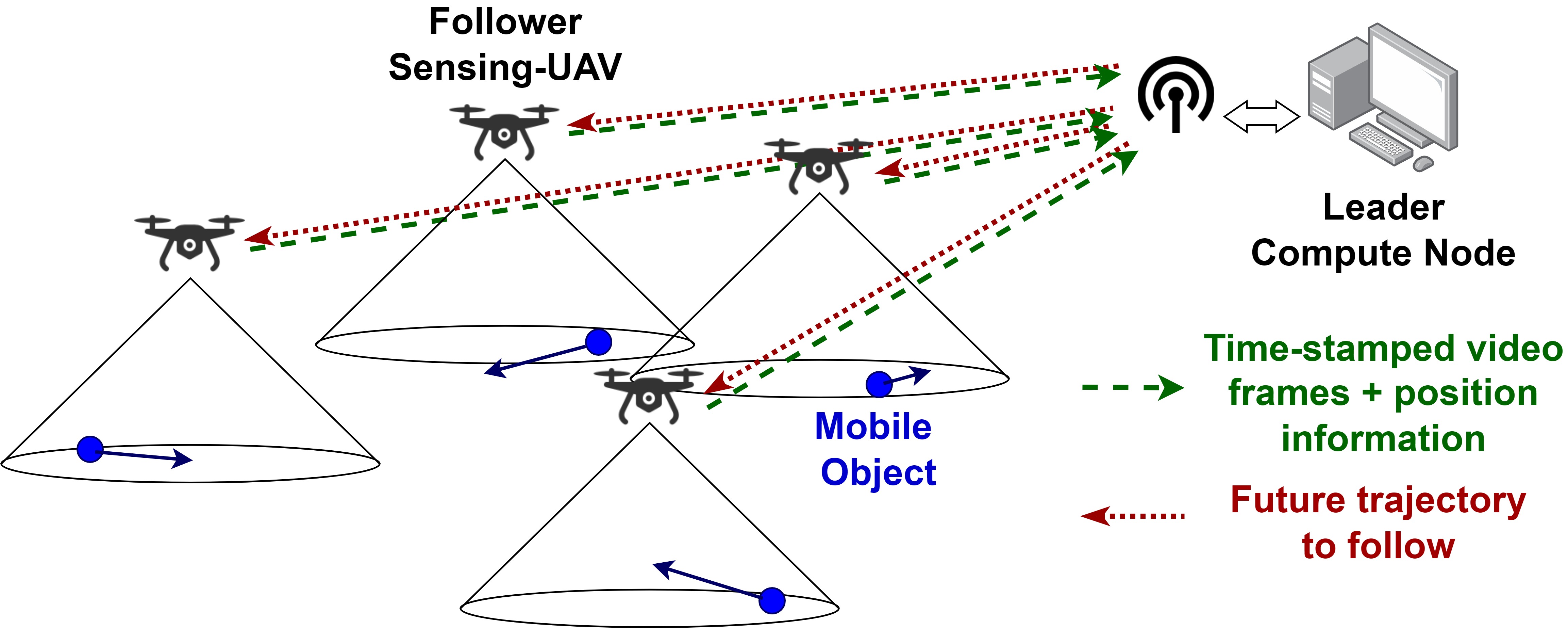}
	\caption{Mobility tracking application implemented using multiple sensing-UAVs and a leader compute node.}
	\label{figure.tracking-model}
\end{figure}

The separation of computing and sensing allows for more scalable system design - with one \emph{leader-node} that has plenty of on-board computational power, and numerous low-cost \emph{sensing-UAVs} that have little computational power but can effectively collect sensor data and communicate over a wireless network. Figure~\ref{figure.tracking-model} illustrates an example of this system design approach. In general, the leader node could be an UAV with a powerful on-board computer such as a Jetson TX2, a compute node located at the wireless edge, or even a cloud server performing high-speed inference and sending back control commands. 

In our specific implementation of the mobility tracking application, the sensing-UAVs capture video of the immediate environment below them and send the captured video frames (without any pre-processing) to the leader compute node. The leader processes the received frames, infers the position of the objects, and sends trajectory updates to the sensing-UAVs via WiFi. 
\emph{The main challenge of this design approach is to manage the limited wireless resources efficiently in order to keep information at the UAVs as fresh as possible.} WiSwarm, an instantiation of our networking middleware, ensures information freshness and scalable tracking performance by carefully controlling the flow of information over the network. 

Next, we describe the different individual components involved in our application - the mobile objects to be tracked, the sensing-UAVs, the leader compute node. We also discuss how WiSwarm is implemented at the sensing-UAVs and the leader compute node. 

\subsection{Mobile Objects}\label{sec.MobileObjects}

We use small autonomous cars equipped with RasPis (3B) as the moving objects whose mobility is tracked by the UAVs. Figure~\ref{fig.sensor-uav-and-car}(b) shows one such car, with the ArUco marker tag on top, which is used for uniquely identifying and tracking the position of the cars by the leader compute node. 


\subsection{Follower Sensing-UAVs}\label{sec.FollowerUAVs}
\begin{figure}[t]
\centering
\subfloat[]
{\includegraphics[width=0.59\columnwidth]{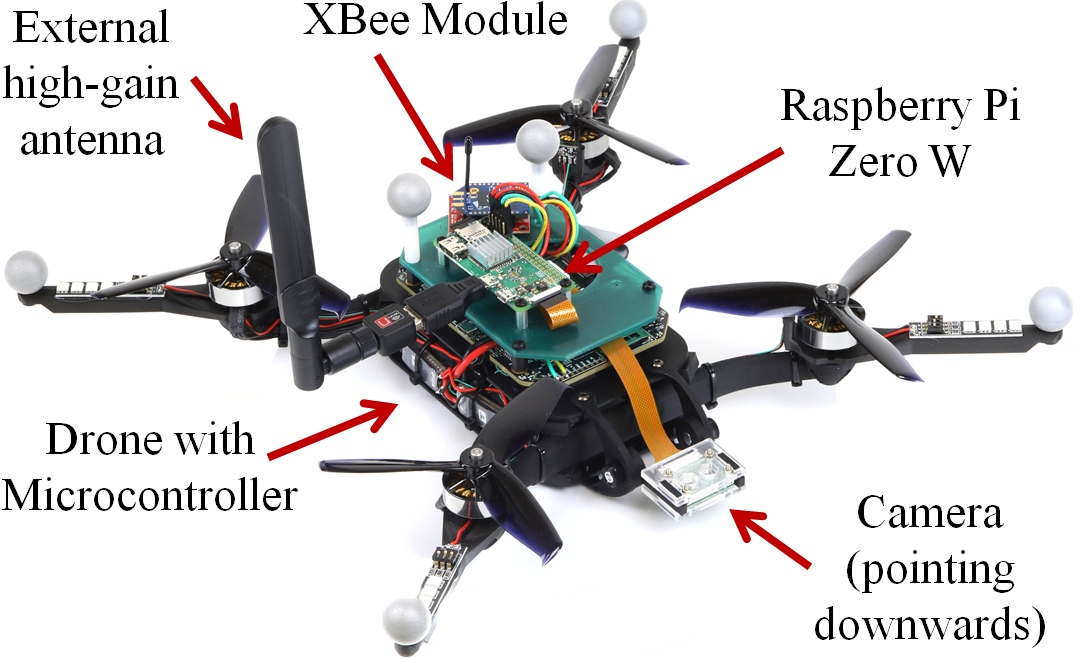}}
\subfloat[]
{\includegraphics[width=0.39\columnwidth]{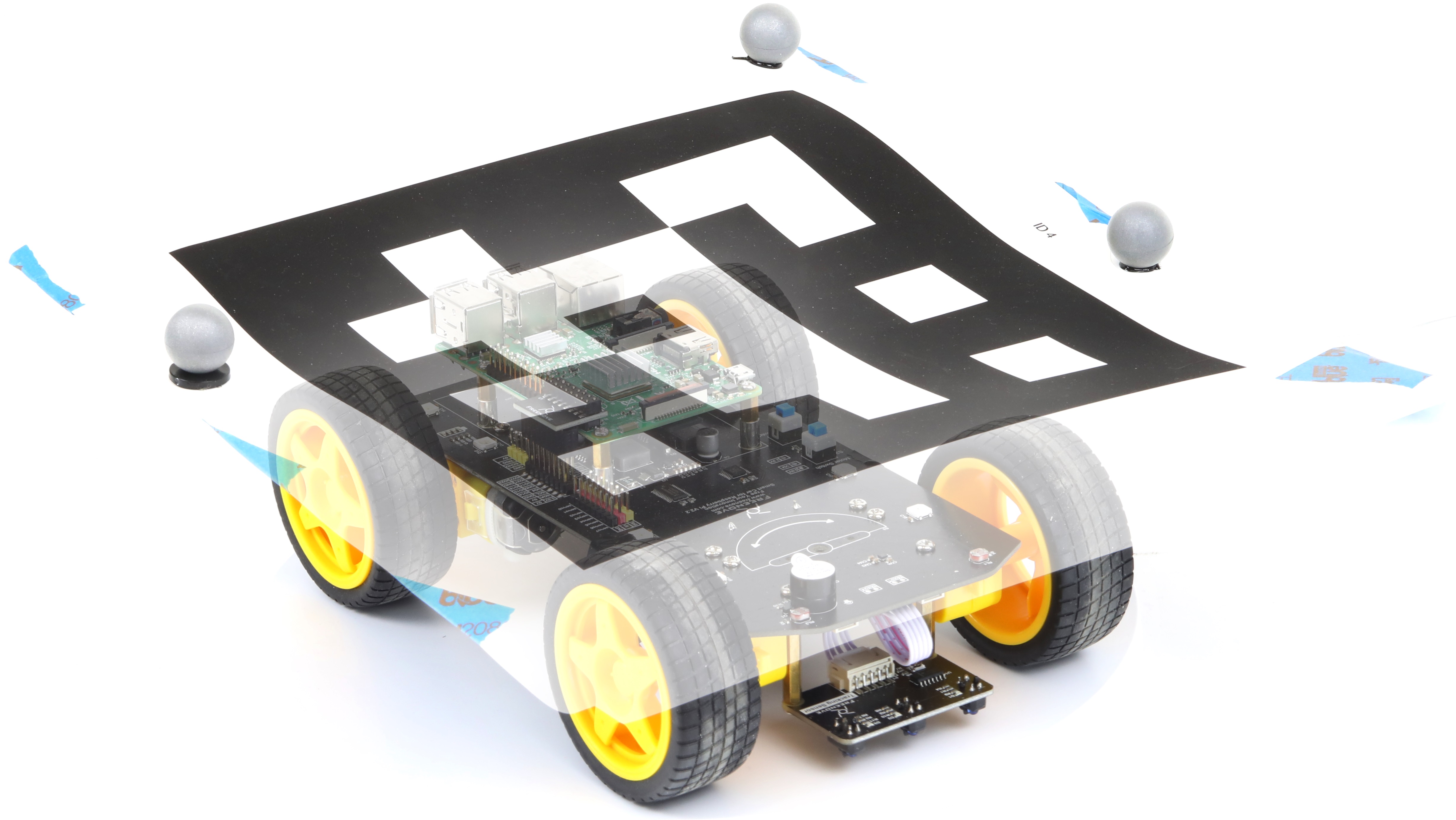}}
\caption{(a) Sensing-UAV. (b) Autonomous car with an identifying ArUco marker on top.} 
\label{fig.sensor-uav-and-car}
\end{figure}
The sensing-UAV consists of two subsystems: a quadcopter drone and a RasPi (Zero W). Figure~\ref{fig.sensor-uav-and-car}(a) shows a sensing-UAV with a RasPi on board the quadcopter drone, along with its sensing and communication peripherals. 

RasPi (Zero Ws) have very little computation capability (1 GHz single-core CPU and 512 MB RAM), but can effectively interact with multiple sensors and also communicate over WiFi. They are also extremely cost-efficient (\$10), making them ideal for use in the sensing-UAVs. Each UAV is also equipped with a micro-controller unit (MCU) that runs state estimation and flight control algorithms. The state estimator combines measurements from an on-board inertial measurement unit (IMU) with global position and orientation measurements.
These global measurements are obtained from a motion capture system and received by an Xbee WiFi module mounted on the UAV.
When motion capture data is not available, the Xbee module can be replaced by an alternative data source, such as a global navigation satellite system (GNSS) receiver.

The RasPi is connected to a camera that captures video of the area below the UAV. Along with each frame, the RasPi also collects the position and orientation at which the frame was collected by asking for this information from the MCU using an asynchronous serial connection. Following the discussion in Sec.~\ref{sec.Queueing}, we know that fresh frames are the most useful for tracking, so we set the queuing discipline at the sensing-UAVs to be LIFO and the buffer size to be such that it can accommodate only one frame at a time. 

The RasPi is connected to the leader compute node over 2.4 GHz WiFi using a high gain (8 dBi) antenna. Whenever the RasPi receives a polling packet, it transmits the most recent update in its LIFO queue to the compute node. 
The RasPi also collects the control information transmitted by the compute node which contains the times and locations (in global coordinates) where the UAV should be in the future in order to track the moving object. The RasPi sends these waypoints over the serial connection to the UAV MCU. The UAV MCU then plans and executes a trajectory that reaches the specified waypoints at the specified future time instants. It does this by interpolating the waypoints to obtain a continuous trajectory that is followed using the flight control algorithm described in \cite{tal2020accurate}. This completes the control loop.

 

\subsection{Leader Compute Node}\label{sec.LeaderNode}
The compute node collects video-frames received from sensing-UAVs in response to polling requests. These video-frames are stored in separate LIFO queues - one for each sensing-UAV. The compute node runs an image processing thread which goes over the queues maintained by WiSwarm in a round-robin manner and processes the received video-frames whenever it finds a non-empty queue.

For each video-frame, the image processing thread attempts to locate the car that the UAV was assigned to track. If the car is found, it uses the relative location of the tag in the frame and the absolute position and orientation at which the frame was captured to compute the global coordinates of the car. The thread also keeps a record of the last known locations of the car. Using the current and previous locations, the image processing thread obtains: (i) the relative velocity between the car and the sensing UAV; and (ii) a list of future waypoints and the time-stamps at which it expects the car to reach these coordinates. In our implementation, we use a simple linear extrapolation scheme to predict future waypoints.

The image processing thread sends the  waypoints and time-stamps to WiSwarm along with information about the relative velocity between the car and the sensing-UAV. WiSwarm uses the relative velocity information to update its  application-defined priority weights 
\begin{equation}
   \textstyle w_i(t) \leftarrow \alpha w_i(t^-) + (1-\alpha) \hat{v}_i(t),
\end{equation}  
where $\hat{v}_i(t)$ is the estimate of relative velocity between the car and the associated sensing UAV, and $\alpha = 0.8$. Since velocity estimates are noisy and car velocities are time-varying, we use an exponential moving average motivated by the adaptive AoI-based scheduling algorithms proposed in \cite{tripathi2021online}. WiSwarm updates link reliabilities $p_i(t)$ by using the number of successful fragment deliveries, as described in Sec.~\ref{sec.MiddlewareDesign}. 

With updated application weights $w_i(t)$ and link reliabilities $p_i(t)$, WiSwarm uses Whittle's Index Policy \eqref{eq:AoI_whittle} to select the sensing-UAVs that need to be scheduled for transmission most urgently. 
Together with the unicast transmission of a polling packet, WiSwarm broadcasts the \emph{most recent} list of future waypoints and time-stamps for every sensing-UAV. This repeated broadcast ensures redundancy in the delivery of control information. 


\section{Evaluation}\label{sec.Evaluation}
In this section, we evaluate the performance of both WiFi and WiSwarm for the mobility tracking application. We perform our experiments in a \emph{dynamic indoor campus space with multiple external sources of interference} such as WiFi base stations, mobile phones, and laptops. Throughout this section when we refer to WiFi, we mean 2.4 GHz WiFi. 

In our evaluation, we consider two experimental setups:
 (i) \textbf{Stationary experiments}, which involve up to fourteen RasPis running an emulated version of the mobility tracking application and sending video-frames to a central Compute Node. These experiments  involved hardware-in-the-loop and allowed us to test a variety of network sizes, update generation rates, scheduling policies, frame resolutions, packet sizes and interference conditions. 
 (ii) \textbf{Flight experiments}, which involve interfacing the RasPis with UAVs and conducting real mobility tracking experiments. These allowed us to test how WiSwarm performs with mobile agents, at longer distances, and in the presence of significant interference. They also illustrate the drawbacks of using WiFi more clearly. 

\textbf{Baseline}. To demonstrate the performance improvement of WiSwarm, we compare it with two baseline WiFi systems, namely WiFi-TCP and WiFi-UDP. Both systems collect video frames from the application layer at a fixed rate, packetize them, store them in FIFO queues, and send these packets over standard WiFi to the Compute Node. TCP uses its congestion control mechanism to adjust the number of packets in flight, while UDP simply forwards packets. In all of our stationary experiments, we found that accommodating the entire video frame within a single UDP packet (i.e., with no fragmentation) was the best choice in terms of tracking error. 

For flight experiments, we consider an \emph{optimized version of WiFi-UDP} as the baseline. Our flight tests showed that mobility tracking with WiFi-UDP and WiFi-TCP with fixed video frame rate (e.g., 50 fps) was not possible for more than a single sensing-UAV. To get mobility tracking to work with two sensing-UAVs, we had to carefully tune the frame generation rate (to 5 fps) and the UDP packet size (to 6 kB per fragment). This is due to the high congestion and unreliability caused by high generation rates and large packets, which caused tracking failures. Further, we also had to tune RTS/CTS thresholds. Despite all of this optimization, WiFi-UDP was only able to enable tracking for \textit{at most two UAVs} at a time, as we show in the discussion on flight experiments.

\subsection{Stationary Experiments}\label{sec.HIL}

\begin{figure}
	\centering
	\includegraphics[width=\linewidth]{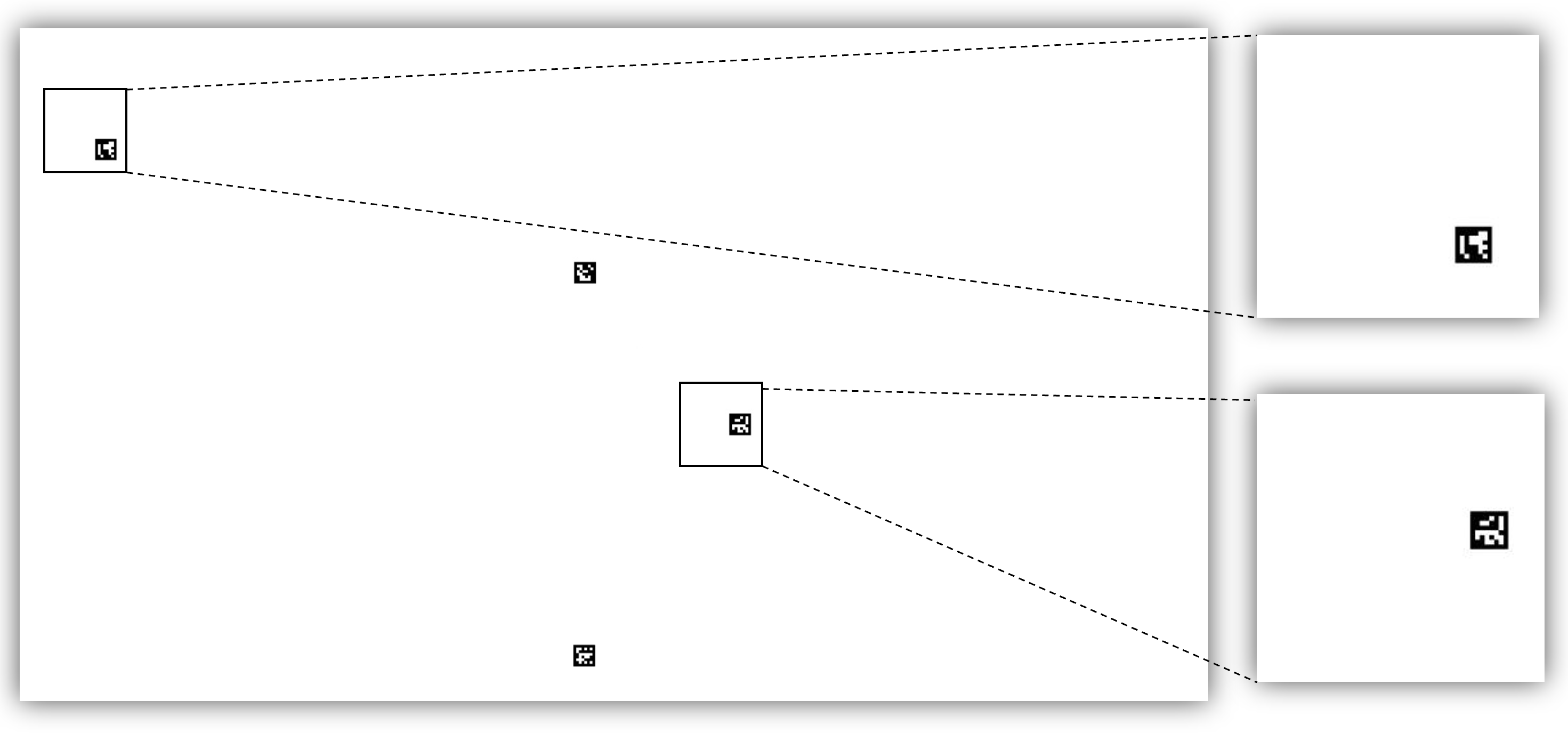}
	\caption{Screenshot from the videos used to simulate car movement during stationary experiments. The tags are programmed to perform random walks with time-varying velocities. The virtual UAVs need to keep track of the tags. On the right, two examples of 224x224 frames sent to the Compute Node by the RasPis based on their current virtual UAV locations.} 
	\label{fig.vid_tags}
\end{figure}


In this section, we discuss the performance improvements of WiSwarm over WiFi for three different metrics: (i) AoI, (ii) throughput, and, most importantly, (iii) tracking error. Each data-point in the following discussion represents 16 minutes worth of experiments, split into 4 batches of 4 minutes each. We calculate the time-average of the performance metric over the entire 4 minutes of each batch and then the mean and standard deviation across batches.
\begin{figure}[t]
\centering
\subfloat[]
{\includegraphics[width=0.5\columnwidth]{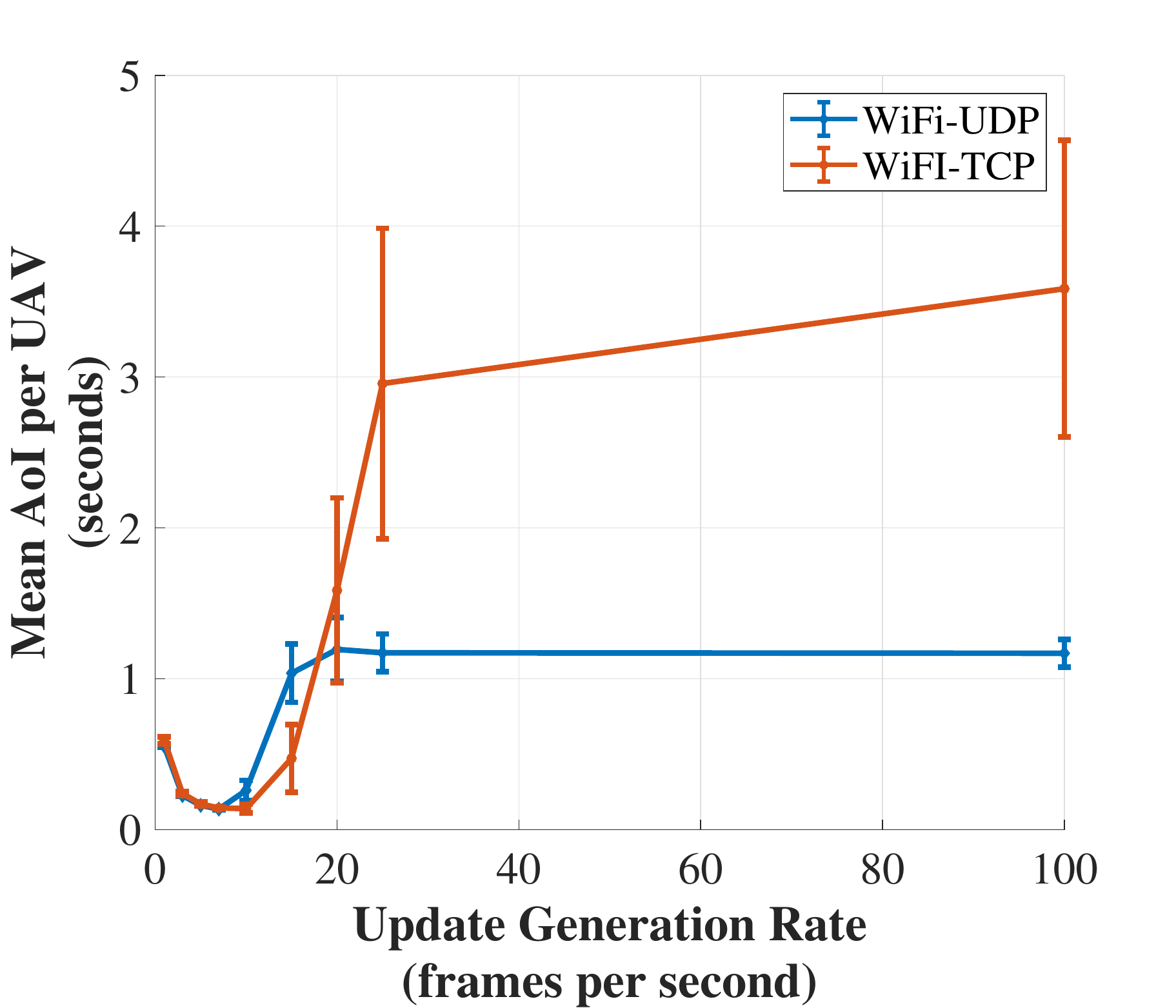}}
\subfloat[]
{\includegraphics[width=0.5\columnwidth]{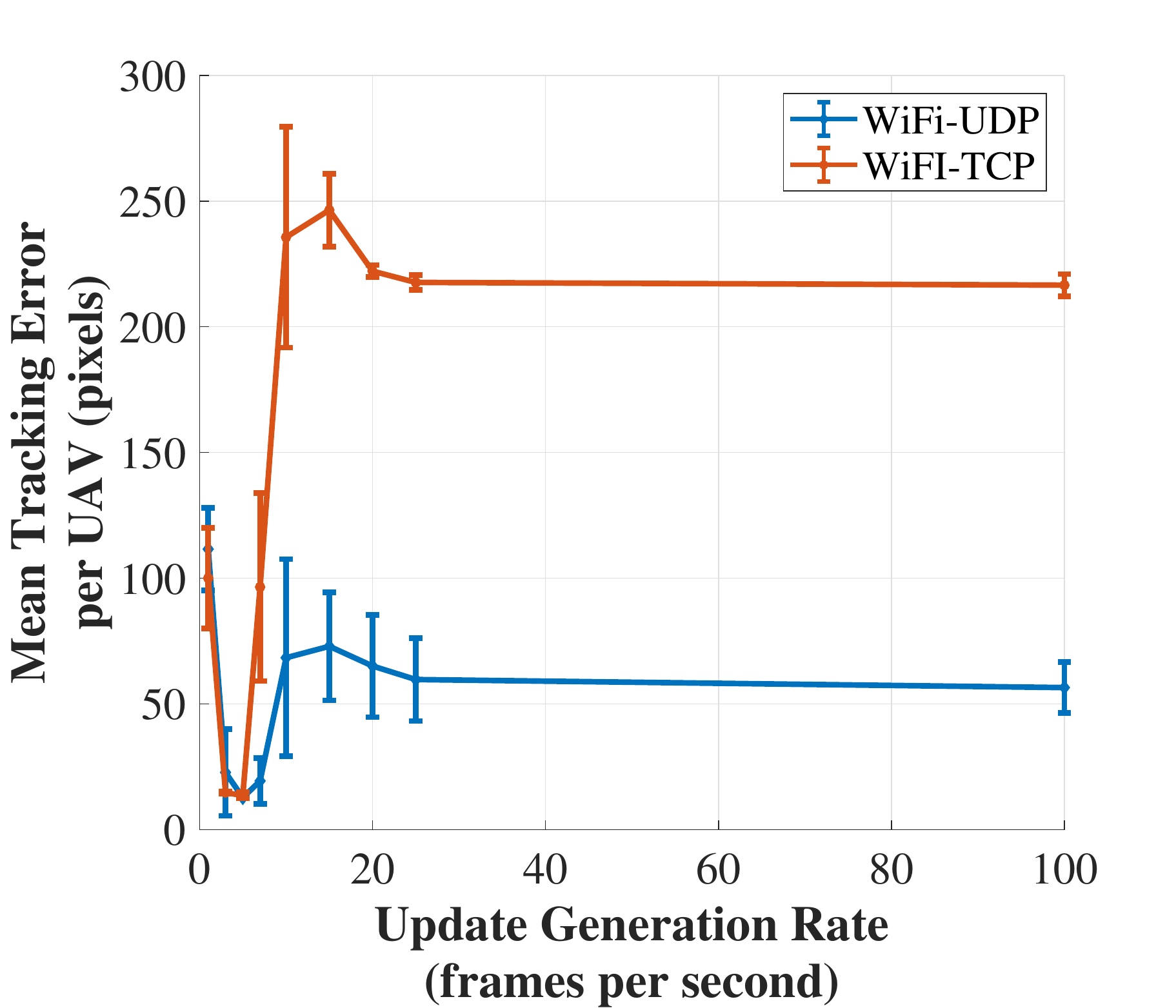}}
\caption{(a) AoI and (b) tracking error of baseline WiFi-TCP and WiFi-UDP plotted against the update generation rate of each of the $N=6$ emulated UAVs.} 
\label{fig.plot_wifi_1}
\end{figure}

\textbf{Experimental Setup}. The experiments involve multiple RasPis running an emulated virtual UAV application. This application does two things. First, each RasPi has a video simulating the movement of cars stored on it. Using this video, the RasPis create cropped frames of size 224x224, based on the current location of the virtual UAV, which capture the local Field-of-View (FoV). These video frames are generated at a specified rate that can be set using the rate control mechanism, and are forwarded to WiFi or WiSwarm for delivery. The frames are stored as unencoded grayscale yuv images (1 byte per pixel), so each video frame is 49 kB in size. Second, the application decodes the control packets received from the Compute Node and updates the virtual UAV's location by moving between control waypoints at a specified speed. Figure~\ref{fig.vid_tags} shows a frame from the video used for simulating movement of the car tags, along with two examples of 224x224 frames that the RasPis send to the Compute Node for processing.   

Figure~\ref{fig.plot_wifi_1} plots the mean AoI and tracking error per UAV for both WiFi-TCP and WiFi-UDP as the frame generation rate at the RasPis increases. This plot is for a system with $6$ transmitting RasPis. Note that lower AoI and lower tracking error are preferred in terms of performance.

We make two important observations from Fig.~\ref{fig.plot_wifi_1}. First, the performance of both WiFi-TCP and WiFi-UDP degrades when the generation rate is high, since the network becomes congested. Second, \textit{WiFi needs optimization of the generation rate at the application layer} to be anywhere close to working in practice. This optimization is challenging since it needs to be at the application layer and also adjust quickly to changes in the traffic load and link reliability, which can vary due to external interference. This is true for both TCP and UDP, i.e. \textit{TCP congestion control was unable to adjust to the optimal rate on its own}.

Next, we compare the performance of WiSwarm with both fixed-rate versions of WiFi and rate-optimized versions of WiFi. We choose the frame generation rates from the set $\{ 1,3,5,7,10,15,20,25,50,100\}$ fps and the number of RasPis from the set $N\in\{ 2,4,6,8,10,12,$ $14\}$. We find the best performing rates for each value of $N$ from the rate set (based on tracking error).

To the best of our knowledge, there are no general purpose systems that can do application layer rate control for a wide variety of real-time applications, so the \textbf{rate-optimized WiFi systems are overly optimistic baselines}. Despite this, WiSwarm achieves significant performance gains over both fixed-rate and optimized rate versions of WiFi-TCP and WiFi-UDP.  

\begin{figure}[t]
\centering
\subfloat[]
{\includegraphics[width=0.5\columnwidth]{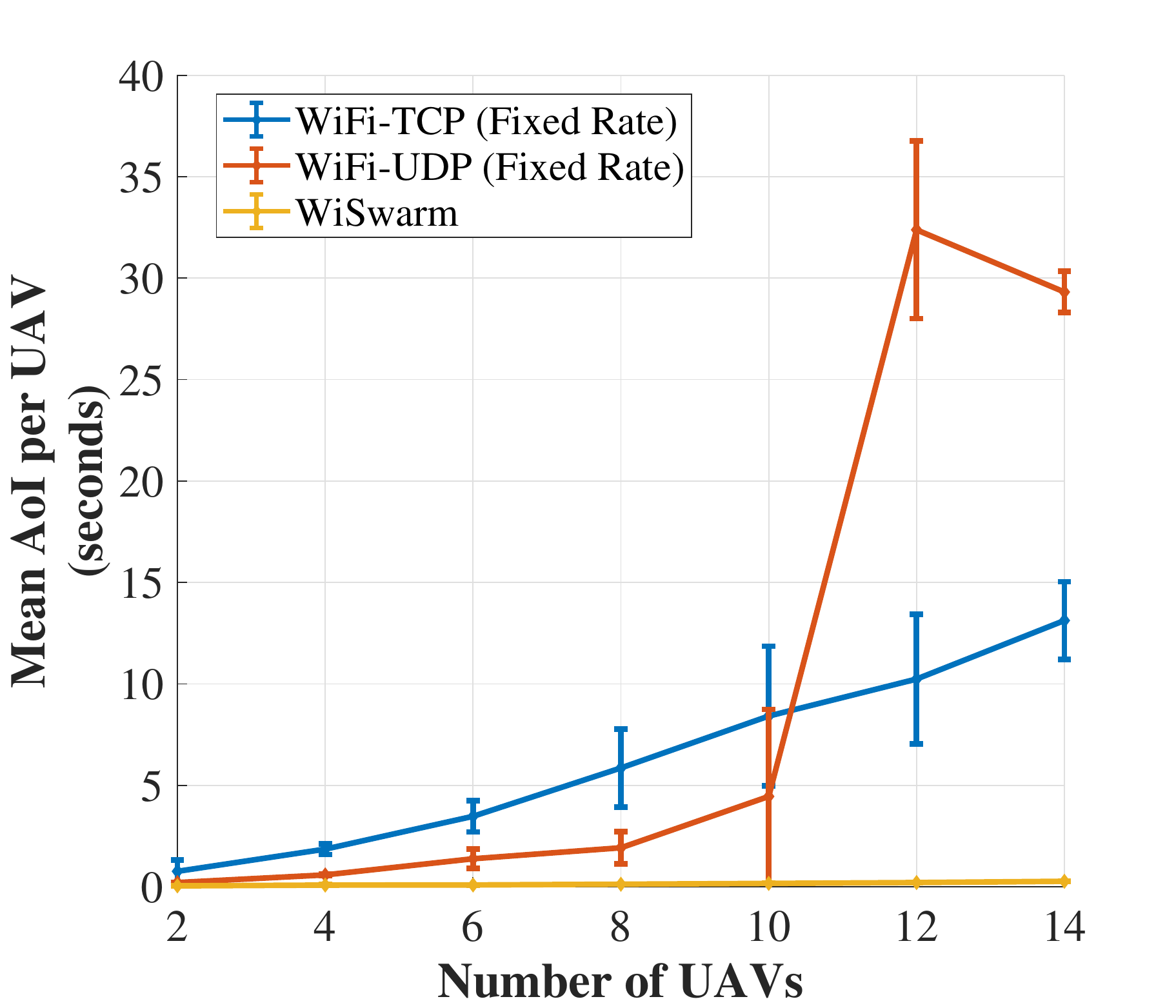}}
\subfloat[]
{\includegraphics[width=0.5\columnwidth]{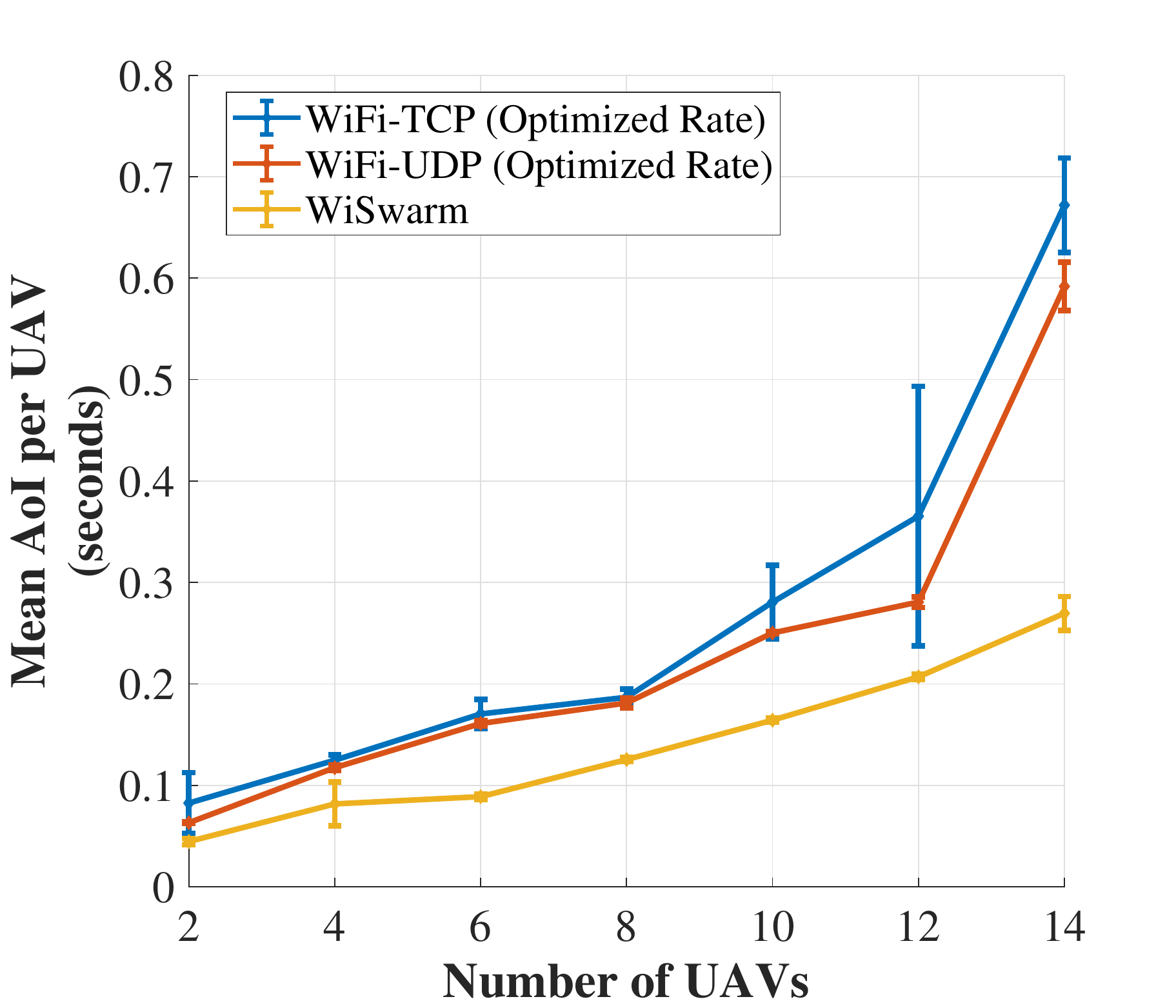}}
\caption{Mean AoI per UAV plotted for (a) fixed-rate (50 fps) and (b) optimized rate WiFi, as well as WiSwarm, as the number of UAVs increases.} 
\label{fig.plot_mean_AoI}
\end{figure}
\begin{figure}[t]
\centering
\subfloat[]
{\includegraphics[width=0.5\columnwidth]{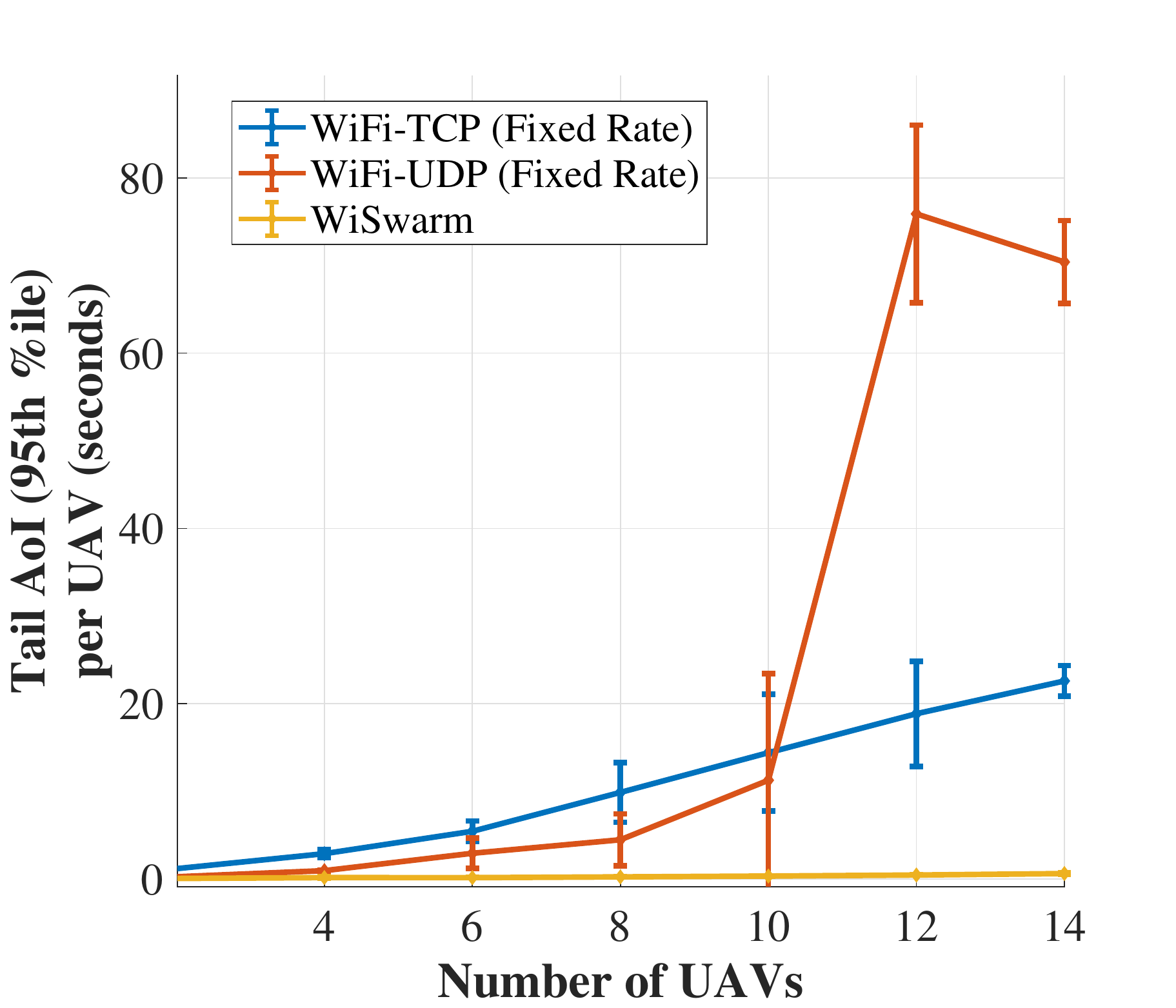}}
\subfloat[]
{\includegraphics[width=0.5\columnwidth]{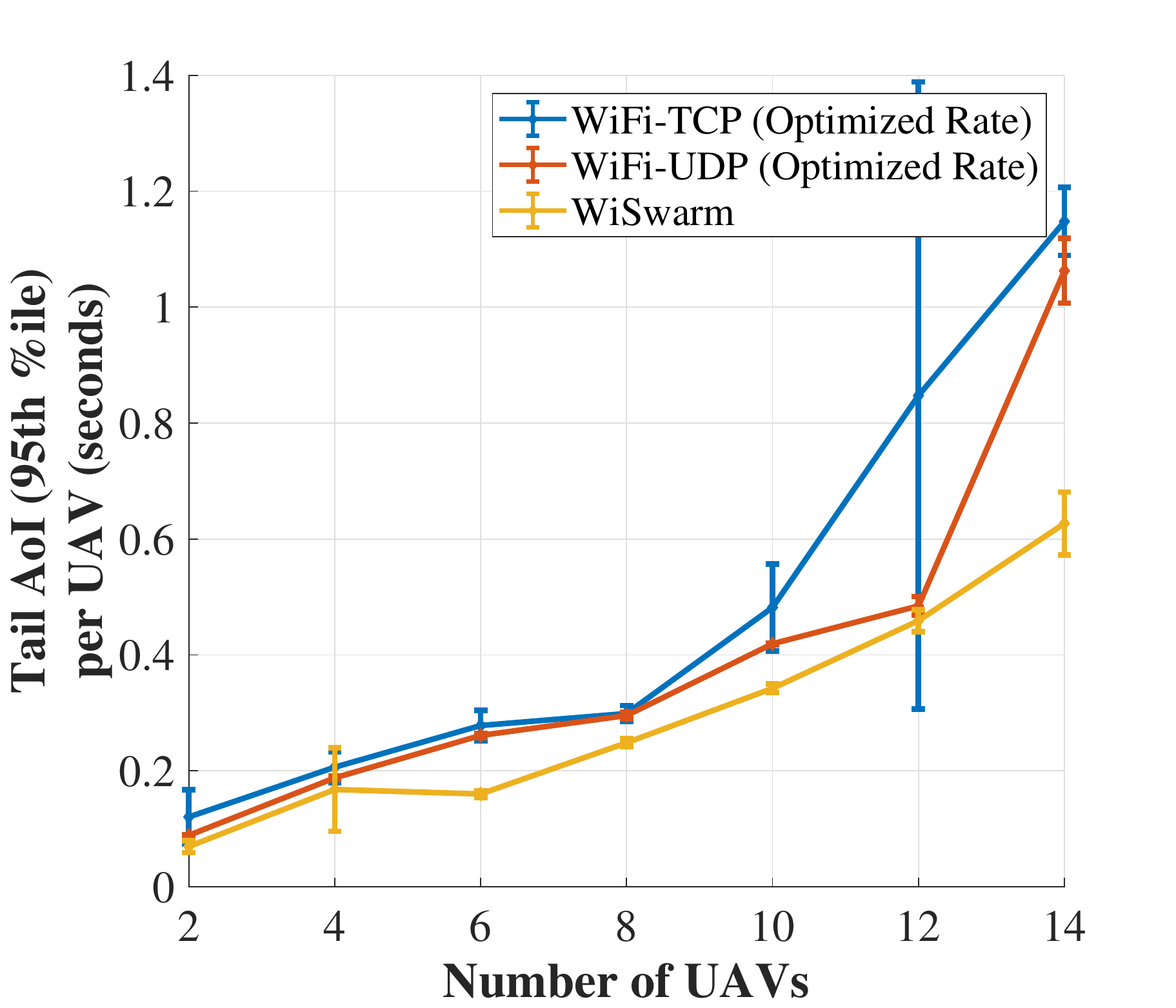}}
\caption{Tail ($95^{th}$ percentile) AoI per UAV plotted for (a) fixed-rate (50 fps) and (b) optimized rate WiFi, as well as WiSwarm, as the number of UAVs increases.} 
\label{fig.plot_tail_AoI}
\end{figure}
\textbf{AoI}. Figure~\ref{fig.plot_mean_AoI} plots the mean AoI per UAV as the system size $N$ increases. More sources in the system means more congestion, more packet collisions (in WiFi) and hence poor performance and scalability. We see this clearly in Fig.~\ref{fig.plot_mean_AoI}(a), where we compare the baseline versions of WiFi-UDP and WiFi-TCP to WiSwarm. The baseline versions of WiFi have fixed update generation rate of 50 fps at each source while WiSwarm uses the maximum generation rate of 100 fps. Mean AoI improves by 16x for $N=8$ and by almost 50x for $N=14$ compared to fixed-rate WiFi. A major cause of the poor performance of WiFi is buildup of FIFO queues once the network becomes congested. Fixed-rate TCP eventually starts outperforming fixed-rate UDP for larger $N$, due to its congestion control mechanism. WiSwarm does not suffer from the congestion problem due to the LIFO queues.

Figure~\ref{fig.plot_mean_AoI}(b) compares rate-optimized versions of WiFi-TCP and WiFi-UDP with WiSwarm. We observe that mean AoI still improves by 1.5x for $N=8$ and 2.2x for $N=14$. While the FIFO queues in WiFi are no longer congested due to careful tuning of the frame generation rates, there are still packet collisions due to the distributed nature of the CSMA protocol and external interference sources. WiSwarm avoids these collisions by centralizing medium access scheduling decisions and prioritizing sources with higher AoI.

Since AoI combines the idea of service regularity with latency, we are also interested in the tail of information freshness. Figure~\ref{fig.plot_tail_AoI} plots the performance of baseline WiFi systems and WiSwarm for the $95th$ percentile of AoI, i.e. the value of AoI which is only exceeded $5\%$ of the time during an entire experiment. We observe very similar gains as mean AoI. For fixed rate, we observe an 18x reduction at $N=8$ and 36x reduction at $N=14$. For rate-optimized, we observe a 1.2x reduction at $N=8$ and a $1.7x$ reduction for $N=14$. Note that the tail AoI is important for our tracking application in addition to mean AoI, since a worse tail suggests a higher probability of the car going out of the UAV's Field-of-View leading to lost tracking. 

\begin{figure}[t]
\centering
\subfloat[]
{\includegraphics[width=0.49\columnwidth]{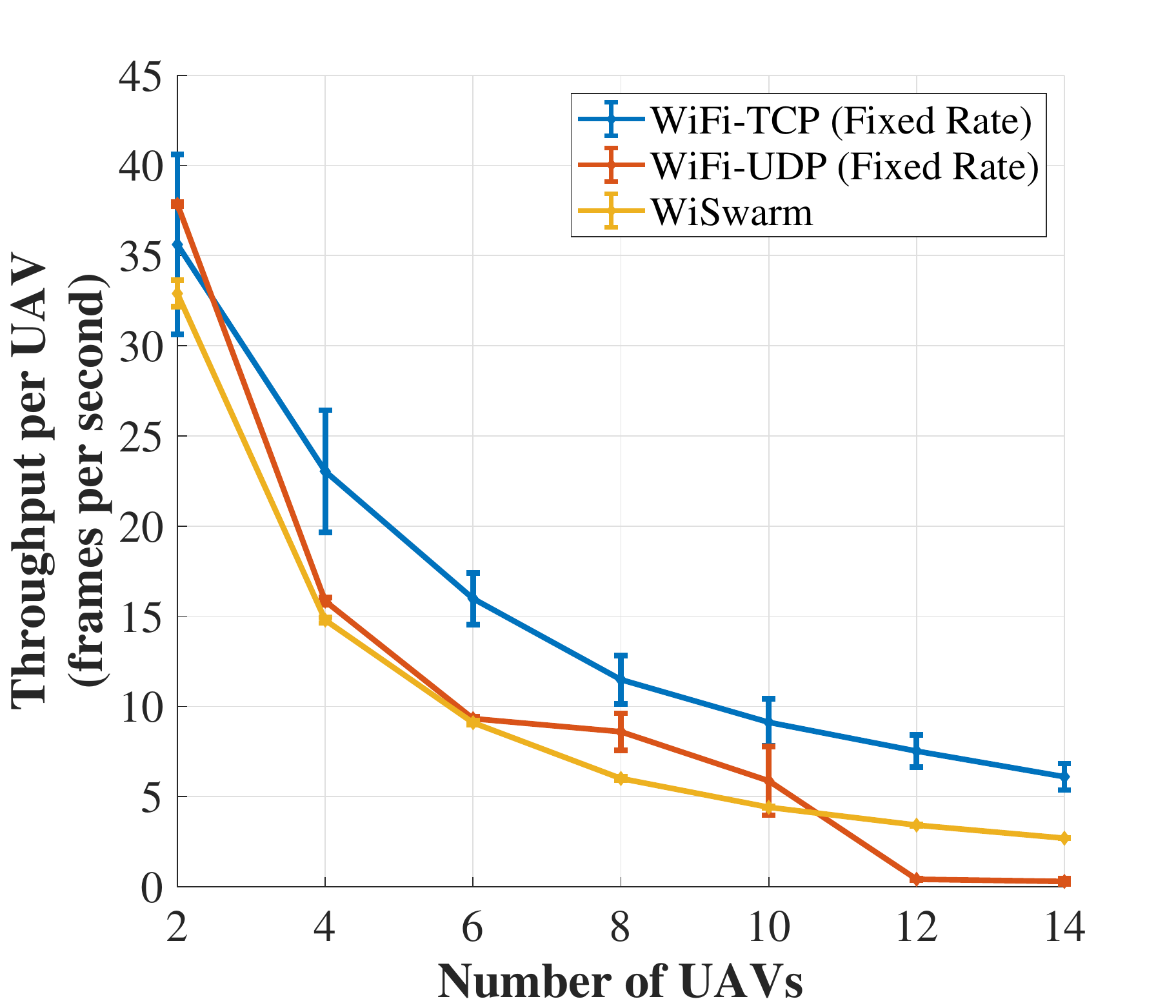}}
\subfloat[]
{\includegraphics[width=0.49\columnwidth]{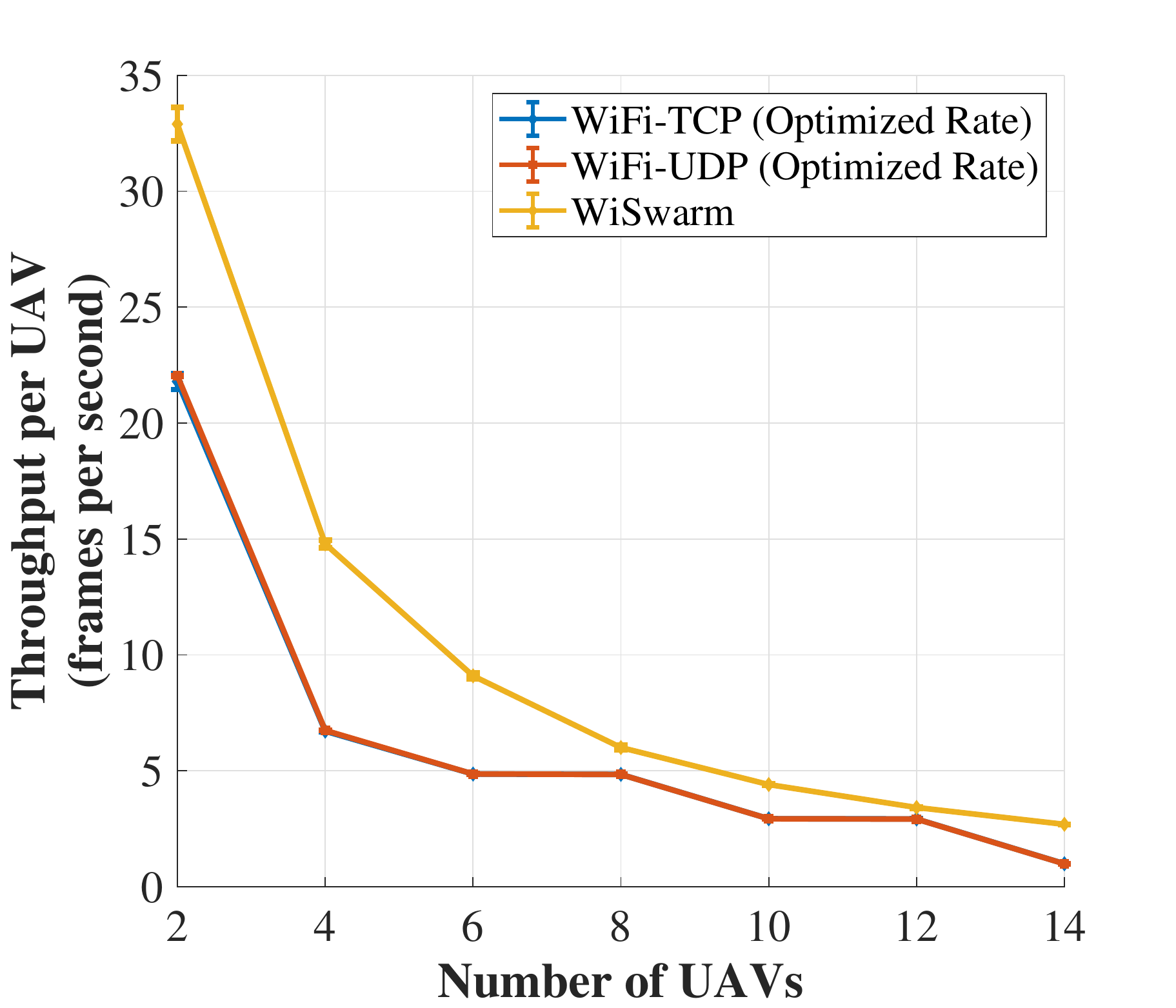}}
\caption{Mean Throughput per UAV plotted for (a) fixed-rate (50 fps) and (b) optimized rate WiFi, as well as WiSwarm, as the number of UAVs increases.} 
\label{fig.plot_throughput}
\end{figure}

\textbf{Throughput}. Figure~\ref{fig.plot_throughput} plots the mean throughput per UAV for each of the considered systems as the number of UAVs increases. From Fig.~\ref{fig.plot_throughput}(a), we observe that both fixed-rate WiFi-TCP and WiFi-UDP have higher per UAV throughput than WiSwarm. However, this doesn't help in getting better AoI (as we saw earlier) or tracking performance (as we will see later). \textbf{This supports the idea that high throughput alone is not sufficient and AoI is the right metric to optimize for in such real-time applications}. For the rate-optimized versions of WiFi, we see a performance improvement in mean throughput per-UAV since WiSwarm can avoid packet collisions and deliver higher rates than the distributed CSMA mechanism while also ensuring lower AoIs. For $N=8$, WiSwarm achieves 1.2x higher throughput and for $N=14$, it achieves 2.7x higher throughput. 

\begin{figure}[t]
\centering
\subfloat[]
{\includegraphics[width=0.5\columnwidth]{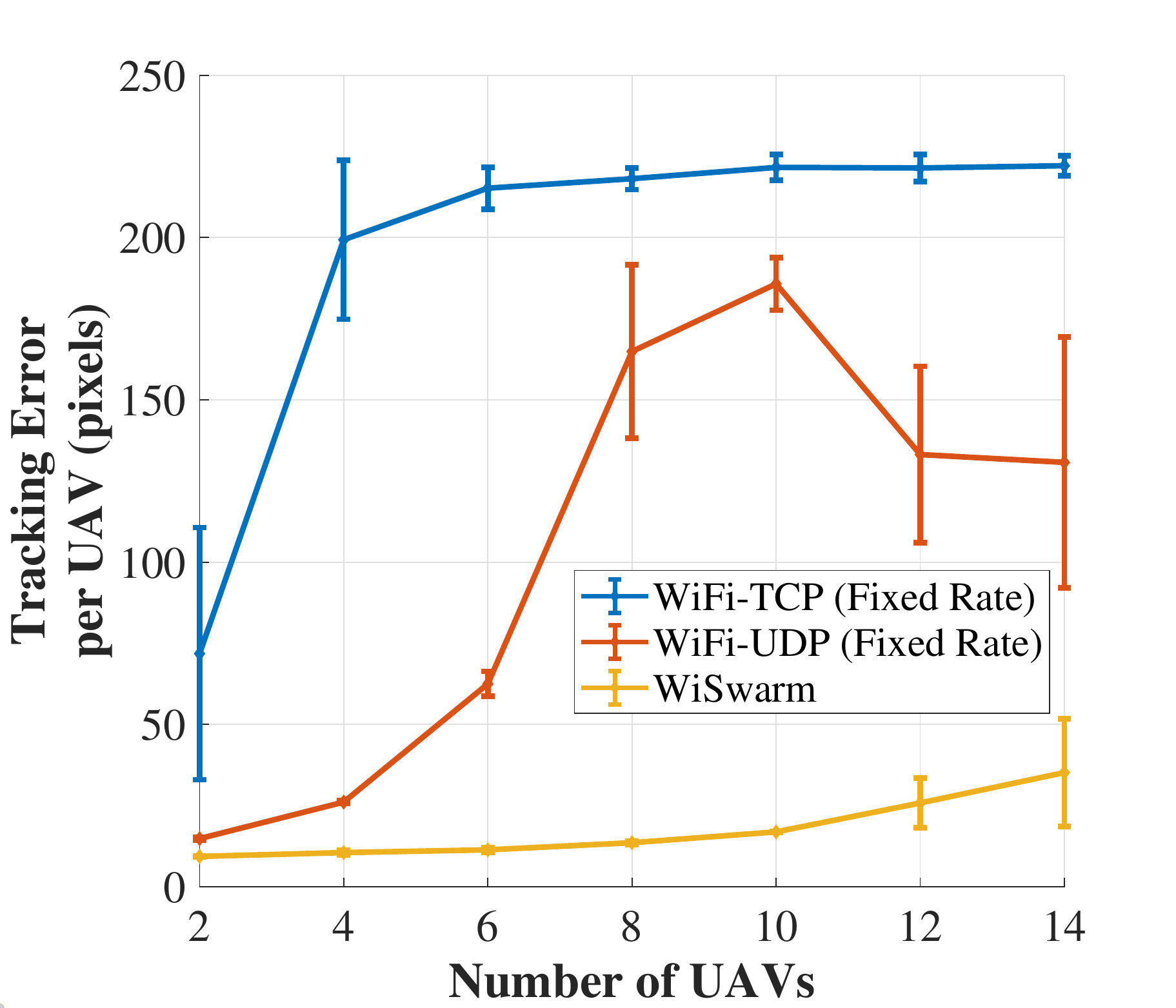}}
\subfloat[]
{\includegraphics[width=0.5\columnwidth]{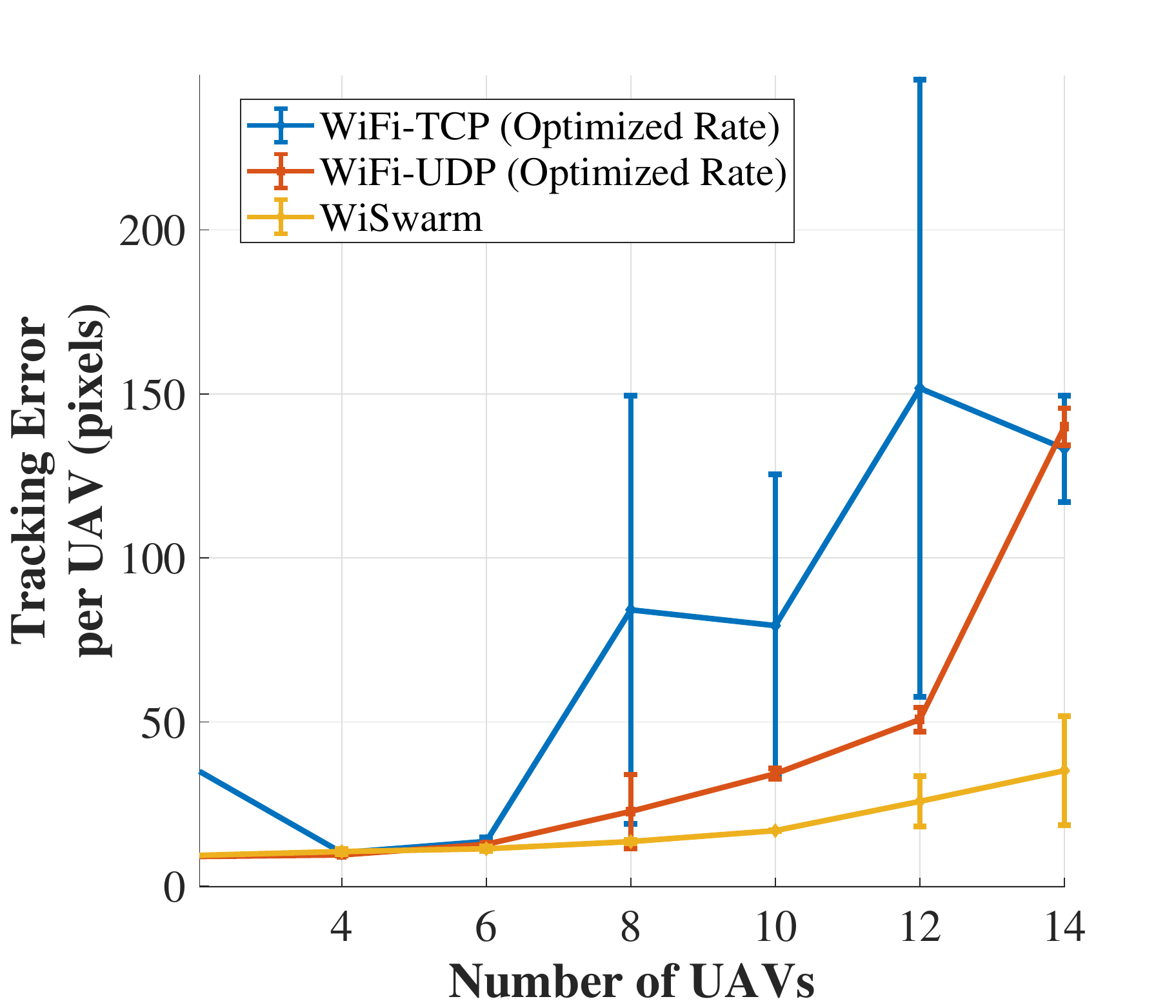}}
\caption{Mean Tracking Error per UAV plotted for (a) fixed-rate (50 fps) and (b) optimized rate WiFi, as well as WiSwarm, as the number of UAVs increases.} 
\label{fig.plot_error}
\end{figure}

\textbf{Tracking Error}. This is where we see how all the pieces of our system design come together to deliver better application performance. Figure~\ref{fig.plot_error} plots the mean tracking error (in pixels) per UAV as the number of UAVs increases, for WiSwarm and WiFi implementations. From Fig.~\ref{fig.plot_error}(a), which shows the fixed-rate baselines, we observe that tracking performance improves by 12x for $N=8$ and 4x for $N=14$. From Fig.~\ref{fig.plot_error}(b), with rate-optimized WiFi versions, we observe that tracking error is reduced by 2x at $N=10$ and 4x at $N=14$ with WiSwarm. We also note that the gap in performance between WiSwarm and the WiFi baselines \textit{increases} with the system size. In other words, the performance of WiFi-TCP and WiFi-UDP degrades much more quickly with $N$ leading to poor scalability.  

\subsection{Flight Experiments}\label{sec.Flight}
\begin{figure}
	\centering
	\includegraphics[width=\linewidth]{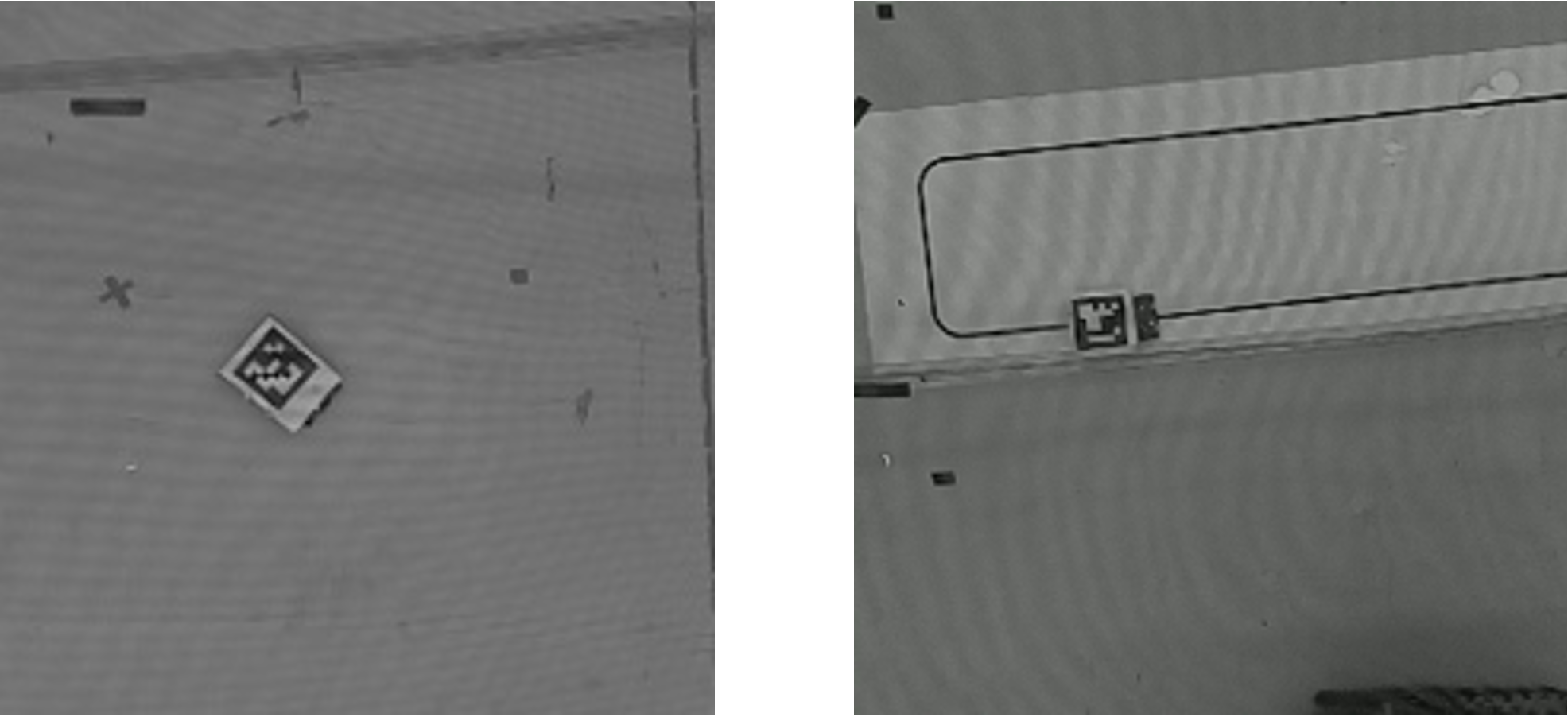}
	\caption{Two Examples of 160x160 grayscale video frames sent by the RasPis during flight experiments.}
	\label{fig.flight_frames}
\end{figure}
While the stationary experiments allowed us to test our system in great detail and provide extensive comparisons, they did not involve implementing the application on real UAVs tracking actual mobile targets in a dynamic environment. Our flight experiments address exactly this setting. Broadly, we will observe that the mobility of UAVs and higher degree of interference leads to worse wireless connectivity and, in turn, more congestion and packet collisions for WiFi. This allows us to bring the robustness of WiSwarm into focus. 
We provide a video describing the setup and results from the flight experiments at \cite{Video}.

\textbf{Experimental Setup}. In the flight tests, we replace the internal antenna of the RasPis with an external high-gain (8 dBi) antenna to improve range and reliability when the UAVs fly. We fly up to 5 UAVs at a time in our experiment space which is roughly 20 meters x 10 meters in size. The mobile objects are autonomous cars with RasPi 3Bs shown in Fig.~\ref{fig.sensor-uav-and-car}(b). We program these cars to move in different polygonal trajectories over time and also stop occasionally at random for a few seconds. These trajectories are unknown to the UAVs and the Compute Node, and the job of the UAVs is to track the cars as closely as possible. Figure~\ref{fig.flight_setup} depicts the setup for an experiment involving 5 UAVs tracking the corresponding cars.

We configure the Pi-Cameras at the UAVs to generate video frames at the maximum possible rate, which is 90 frames per second. For WiSwarm, we utilize this full rate, while for WiFi, we choose the optimized rate by using rate control. The video frames are 160x160 unencoded grayscale images in the yuv format (1 byte per pixel), with a total size of 25 kB per frame. 
Figure~\ref{fig.flight_frames} shows two examples of frames sent to the Compute Node by RasPis from the flying UAVs during different experiments.

The sensing-UAVs implement a controller that requires knowledge of their own global position and orientation to be able to plan desired trajectories. A Motion Capture (MoCap) system provides this information to the UAVs (also via 2.4 GHz WiFi). 
These MoCap messages are sent to the UAVs in UDP messages at 30 messages/second and each message contains timestamp, position, and orientation of a single vehicle in 45 bytes. So the MoCap network usage is approximately 1.3 kB/s (or 11 kb/s) per UAV. Importantly, the MoCap system runs completely independently from the WiSwarm and WiFi systems and causes a low level of persistent interference in the channel. Thus, results from our flight experiments are a good measure of robustness of WiSwarm and WiFi to external interference. 



\begin{figure}
	\centering
	\includegraphics[width=\linewidth]{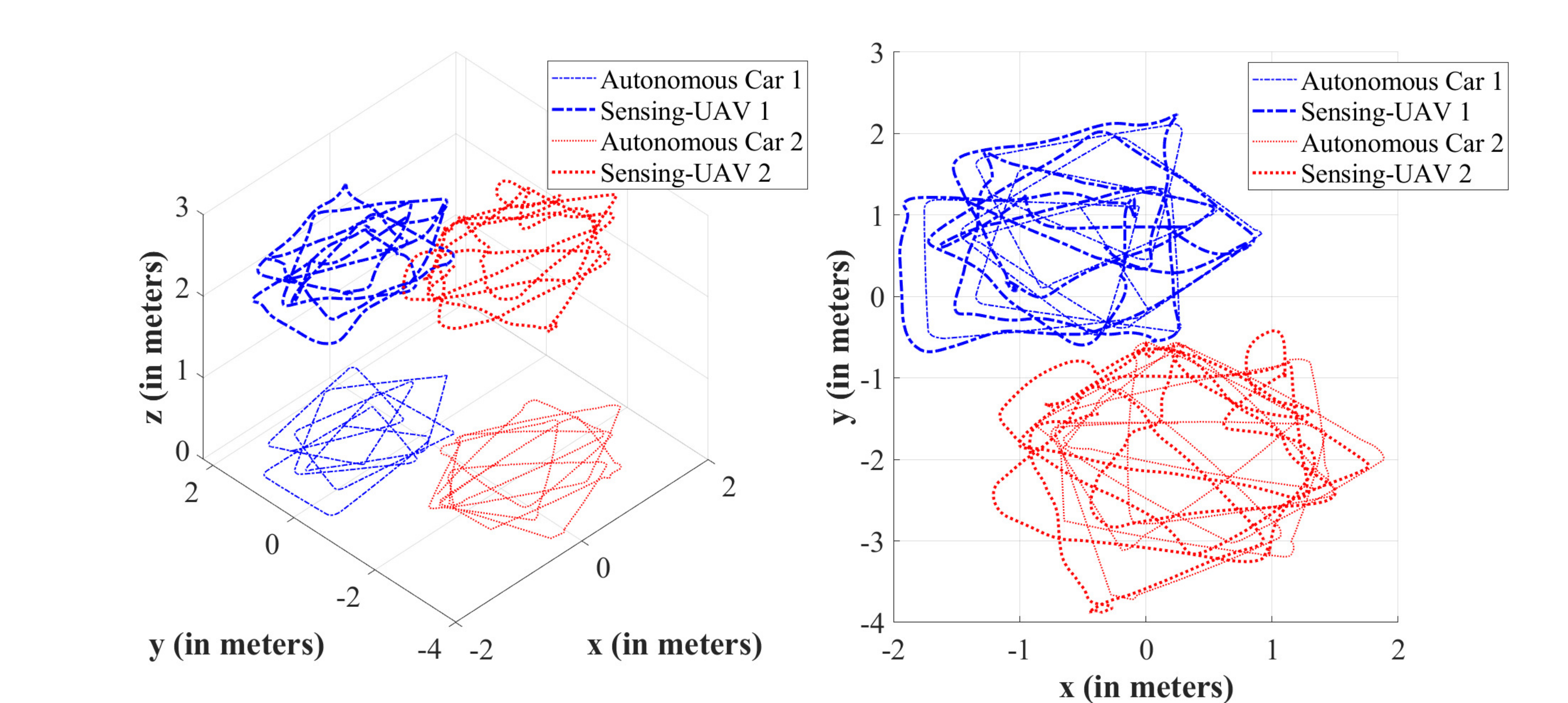}
	\caption{Coordinates of sensing-UAVs and target cars in 2-D and 3-D, for a two drone flight experiment running WiSwarm.}
	\label{fig.coord_wiswarm}
\end{figure}

\begin{figure}
	\centering
	\includegraphics[width=\linewidth]{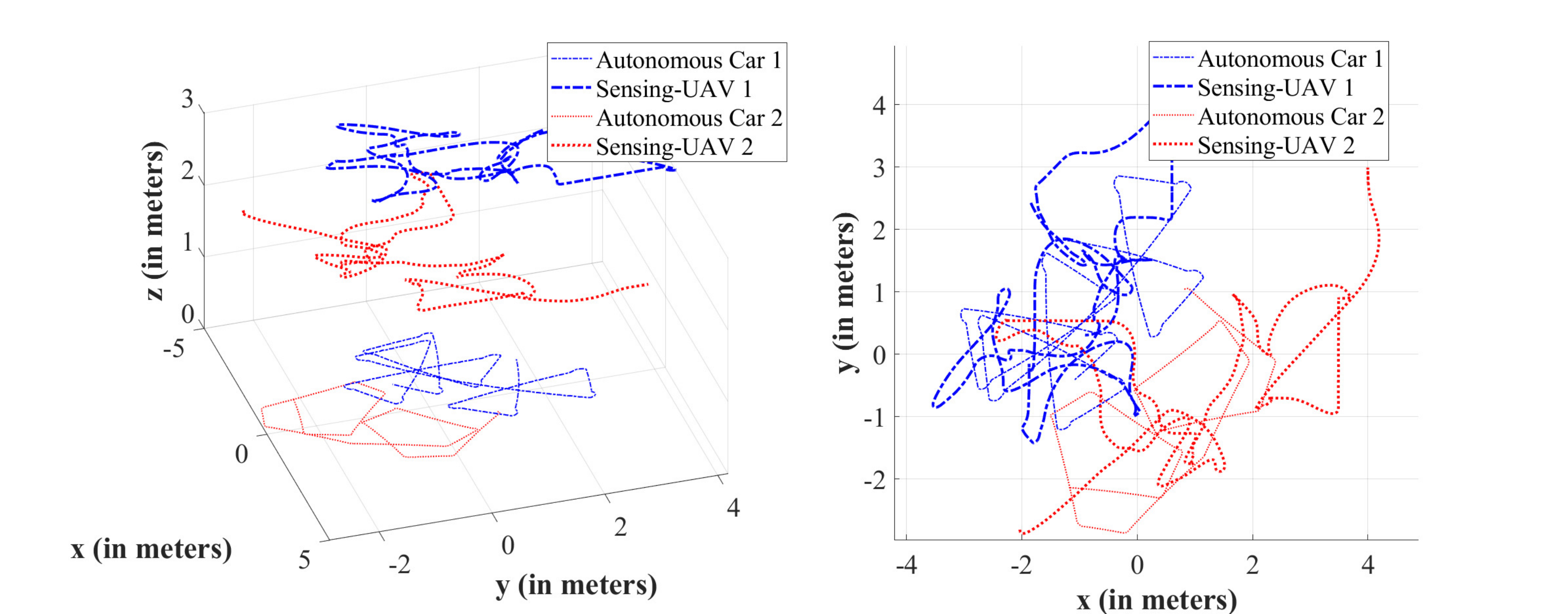}
	\caption{Coordinates of sensing-UAVs and target cars in 2-D and 3-D, for a two drone flight experiment running optimized WiFi-UDP.}
	\label{fig.coord_wiFi}
\end{figure}

\begin{figure}[t]
\centering
\subfloat[]
{\includegraphics[width=0.49\columnwidth]{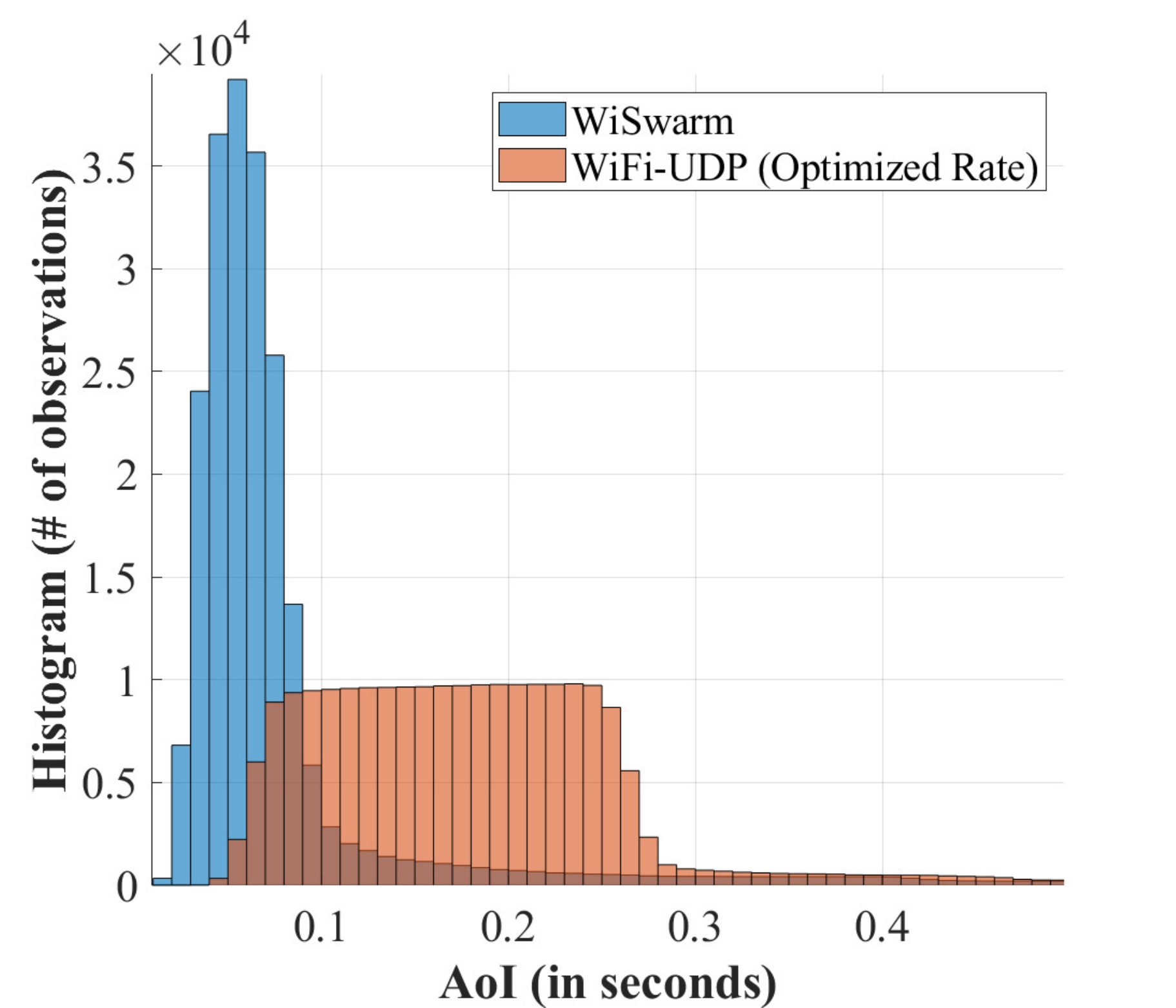}}
\subfloat[]
{\includegraphics[width=0.49\columnwidth]{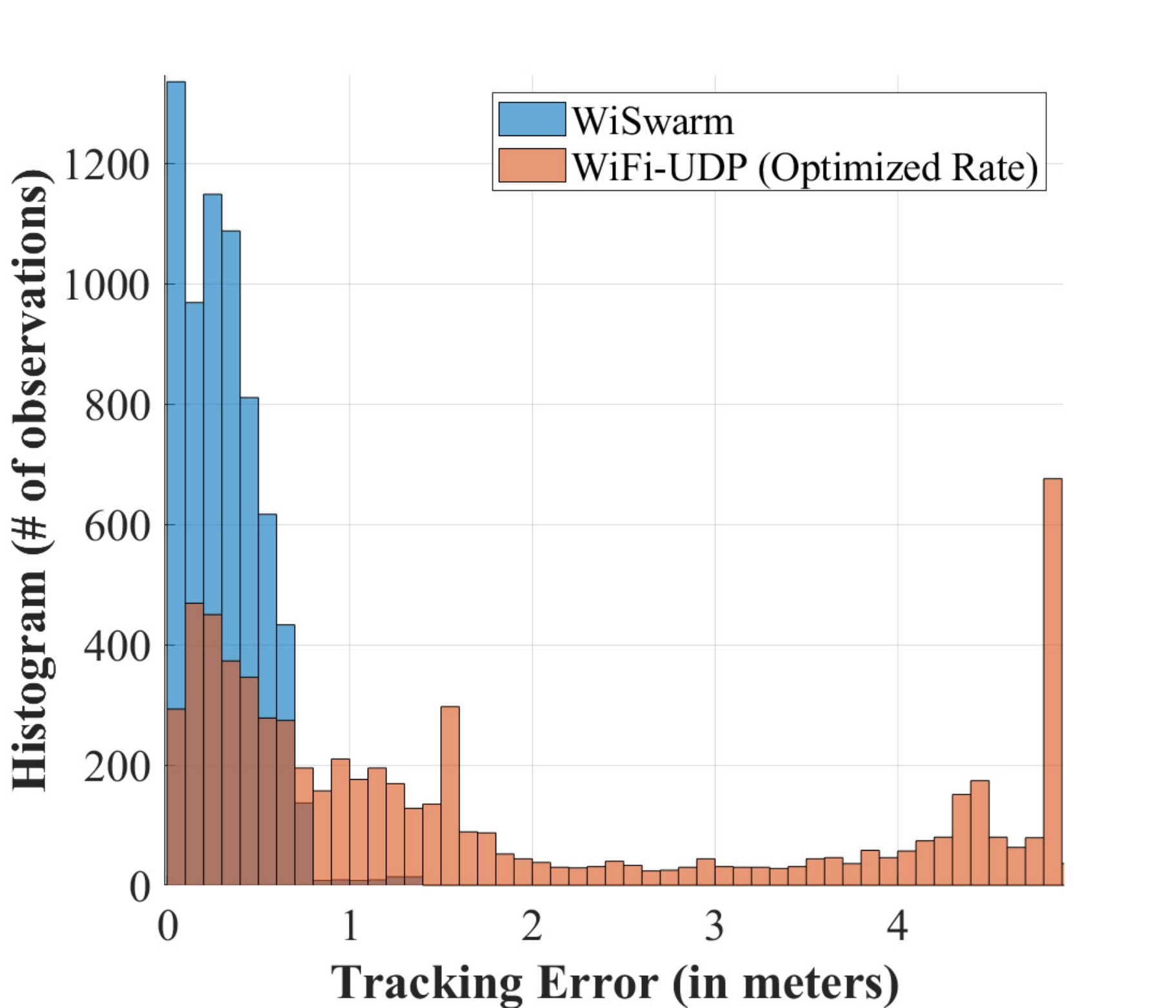}}
\caption{Histograms of (a) AoI and (b) tracking error for flight experiments with two UAVs, comparing WiSwarm with WiFi.} 
\label{fig.flight_histogram}
\end{figure}

\textbf{Results}. Figure~\ref{fig.coord_wiswarm} plots the coordinates of the sensing-UAVs and the target cars over time, for a two drone WiSwarm experiment, in both 2-D and 3-D. Similarly, Fig.~\ref{fig.coord_wiFi} plots the coordinates of the sensing-UAVs and the target cars over time, for a two drone WiFi experiment. It is easy to see that WiSwarm allows for far better tracking than WiFi even for just two UAVs. This is further supported by the histograms of AoI and tracking error plotted in Fig.~\ref{fig.flight_histogram}. The lower tracking error for WiSwarm is \textit{due to the fact that it can achieve lower AoI}, and hence deliver fresher information.

We summarize the results of all of our flight experiments in Tables \ref{table:error} and \ref{table:AoI}. We average over 4 minutes of flight data for each experiment. Our main observation is as follows: \textbf{while WiFi allows tracking for up to two UAVs at a time, WiSwarm can easily allow tracking for up to five UAVs at a time}. In fact, when there are more than two sources in the system, WiFi is unable to deliver more than a handful of packets and essentially no UAV control is possible. The main reason for this is the high level of packet collisions for WiFi. WiSwarm is relatively robust to the unreliable wireless channels, interference and mobility issues encountered in flight experiments, due to our scheduler design that avoids packet collisions and prioritizes AoI.

\begin{table}[h!]
\centering
\begin{tabular}{|c|c|c|c|c|c|} 
 \hline
 Number of Drones & 1 & 2 & 3 & 4 & 5 \\ [0.3ex] 
 \hline\hline
 WiFi-UDP (Optimized) & 0.43 & 1.85 & - & - & - \\ 
 \hline
 WiSwarm & 0.39 & 0.30 & 0.39 & 0.35 & 0.36\\[0.3ex] 
 \hline
\end{tabular}
\caption{Average tracking error per sensing-UAV (in meters).}
\label{table:error}
\end{table}

\begin{table}[h!]
\centering
\begin{tabular}{|c|c|c|c|c|c|} 
 \hline
 Number of Drones & 1 & 2 & 3 & 4 & 5 \\ [0.3ex] 
 \hline\hline
 WiFi-UDP (Optimized) & 0.10 & 0.19 & - & - & - \\ 
 \hline
 WiSwarm & 0.08 & 0.09 & 0.11 & 0.12 & 0.16\\[0.3ex] 
 \hline
\end{tabular}
\caption{Average AoI per sensing-UAV (in seconds).}
\label{table:AoI}
\end{table}
\section{Conclusion}\label{sec.Conclusion}
In this paper, we propose an AoI-based networking middleware that enables the customization of WiFi networks to the needs of time-sensitive applications that rely on multi-agent systems. By controlling the storage and flow of information in the underlying WiFi network, the middleware can prevent packet collisions and dynamically prioritize transmissions aiming to optimize information freshness. The middleware is implemented at the application layer, facilitating customization and integration to existing systems 
To demonstrate the benefits of our middleware, we implement a mobility tracking application using a swarm of sensing-UAVs communicating with a central controller via WiFi. Our experimental results show that our middleware can improve information freshness and, as a result, tracking accuracy by \emph{more than one order of magnitude} when compared to an equivalent system that uses plain WiFi. Our flight tests also show that the middleware improves scalability of the mobility tracking application. Interesting extensions of this work include consideration of a distributed middleware architecture. 

\bibliographystyle{ieeetr}
\bibliography{bibliography_2}

\begin{thebibliography}{10}

\bibitem{KivaWiFi}
P.~Valerio, ``Amazon robotics: {IoT} in the warehouse.'' online:
  https://www.informationweek.com/strategic-cio/amazon-robotics-iot-in-the-warehouse/d/d-id/1322366,
  2015.

\bibitem{VehicleToInfrastructureDSRC}
J.~Gozalvez, M.~Sepulcre, and R.~Bauza, ``{IEEE} 802.11p vehicle to
  infrastructure communications in urban environments,'' {\em IEEE Commun.
  Magazine}, vol.~50, no.~5, pp.~176--183, 2012.

\bibitem{AppLevelDSRC}
M.~Klapez, C.~A. Grazia, and M.~Casoni, ``Application-level performance of
  {IEEE} 802.11p in safety-related {V2X} field trials,'' {\em IEEE Internet of
  Things Journal}, vol.~7, no.~5, pp.~3850--3860, 2020.

\bibitem{NYCDOT}
{New York City Department of Transportation}, ``{NYC} connected vehicle
  project: For safer transportation.'' online: https://cvp.nyc/, 2022.

\bibitem{CERBERUS}
M.~Tranzatto, F.~Mascarich, L.~Bernreiter, C.~Godinho, M.~Camurri, S.~Khattak,
  T.~Dang, V.~Reijgwart, J.~Loje, D.~Wisth, S.~Zimmermann, H.~Nguyen, M.~Fehr,
  L.~Solanka, R.~Buchanan, M.~Bjelonic, N.~Khedekar, M.~Valceschini,
  F.~Jenelten, M.~Dharmadhikari, T.~Homberger, P.~D. Petris, L.~Wellhausen,
  M.~Kulkarni, T.~Miki, S.~Hirsch, M.~Montenegro, C.~Papachristos, F.~Tresoldi,
  J.~Carius, G.~Valsecchi, J.~Lee, K.~Meyer, X.~Wu, J.~Nieto, A.~Smith,
  M.~Hutter, R.~Siegwart, M.~Mueller, M.~Fallon, and K.~Alexis, ``{CERBERUS}:
  Autonomous legged and aerial robotic exploration in the tunnel and urban
  circuits of the {DARPA} subterranean challenge,'' {\em Field Robotics}, 2021.

\bibitem{SAR}
J.~Cacace, A.~Finzi, V.~Lippiello, M.~Furci, N.~Mimmo, and L.~Marconi, ``A
  control architecture for multiple drones operated via multimodal interaction
  in search \& rescue mission,'' in {\em Proc. IEEE Int. Symp. Safety,
  Security, and Rescue Robotics (SSRR)}, 2016.

\bibitem{DistributedRobotFormation}
J.~Alonso-Mora, E.~Montijano, T.~Nägeli, O.~Hilliges, M.~Schwager, and D.~Rus,
  ``Distributed multi-robot formation control in dynamic environments,'' {\em
  Autonomous Robots}, vol.~43, p.~1079–1100, 2019.

\bibitem{CooperativeSLAM}
N.~Mahdoui, V.~Frémont, and E.~Natalizio, ``Communicating multi-{UAV} system
  for cooperative {SLAM}-based exploration,'' {\em Journal of Intelligent \&
  Robotic Systems}, vol.~98, p.~325–343, 2020.

\bibitem{ASTRO}
R.~Petrolo, Y.~Lin, and E.~Knightly, ``{ASTRO}: Autonomous, sensing, and
  tetherless networked drones,'' in {\em Proc. ACM MobiSys – DroNet Wkshp.},
  2018.

\bibitem{DroneCinema}
T.~N\"{a}geli, L.~Meier, A.~Domahidi, J.~Alonso-Mora, and O.~Hilliges,
  ``Real-time planning for automated multi-view drone cinematography,'' {\em
  ACM Trans. Graphics}, vol.~36, no.~4, 2017.

\bibitem{DOOR-SLAM}
P.-Y. Lajoie, B.~Ramtoula, Y.~Chang, L.~Carlone, and G.~Beltrame,
  ``{DOOR-SLAM:} distributed, online, and outlier resilient slam for robotic
  teams,'' {\em IEEE Robot. Autom. Letters}, vol.~5, no.~2, pp.~1656--1663,
  2020.

\bibitem{BeeCluster}
S.~He, F.~Bastani, A.~Balasingam, K.~Gopalakrishna, Z.~Jiang, M.~Alizadeh,
  H.~Balakrishnan, M.~Cafarella, T.~Kraska, and S.~Madden, ``Beecluster: Drone
  orchestration via predictive optimization,'' in {\em Proc. ACM MobiSys},
  2020.

\bibitem{MultiRobotSlam}
M.~T. Lázaro, L.~M. Paz, P.~Pinies, J.~A. Castellanos, and G.~Grisetti,
  ``Multi-robot {SLAM} using condensed measurements,'' in {\em Proc. IEEE/RSJ
  IROS}, 2013.

\bibitem{MultiRobotMapping}
L.~Matignon and O.~Simonin, ``Multi-robot simultaneous coverage and mapping of
  complex scene - comparison of different strategies,'' in {\em Proc. ACM
  AAMAS}, 2018.

\bibitem{hu2020hivemind}
J.~Hu, A.~Bruno, B.~Ritchken, B.~Jackson, M.~Espinosa, A.~Shah, and
  C.~Delimitrou, ``Hivemind: A scalable and serverless coordination control
  platform for {UAV} swarms,'' {\em arXiv preprint arXiv:2002.01419}, 2020.

\bibitem{chinchali2021network}
S.~Chinchali, A.~Sharma, J.~Harrison, A.~Elhafsi, D.~Kang, E.~Pergament,
  E.~Cidon, S.~Katti, and M.~Pavone, ``Network offloading policies for cloud
  robotics: a learning-based approach,'' {\em Autonomous Robots}, vol.~45,
  no.~7, pp.~997--1012, 2021.

\bibitem{Video}
``Wiswarm: Age-of-information-based wireless networking for collaborative teams
  of uavs.''
  \url{https://www.dropbox.com/s/bu39jkujassyagr/WiSwarm_INFOCOM_2023.mp4?dl=0},
  2022.
\newblock Video Attachment.

\bibitem{kaul2012real}
S.~Kaul, R.~Yates, and M.~Gruteser, ``Real-time status: How often should one
  update?,'' in {\em Proc. IEEE INFOCOM}, 2012.

\bibitem{yates2021age}
R.~D. Yates, Y.~Sun, D.~R. Brown, S.~K. Kaul, E.~Modiano, and S.~Ulukus, ``Age
  of information: An introduction and survey,'' {\em IEEE J. Sel. Areas
  Commun}, vol.~39, no.~5, pp.~1183--1210, 2021.

\bibitem{kosta2017age}
A.~Kosta, N.~Pappas, V.~Angelakis, {\em et~al.}, ``Age of information: A new
  concept, metric, and tool,'' {\em Foundations and Trends in Networking},
  vol.~12, no.~3, pp.~162--259, 2017.

\bibitem{sun2019age_book}
Y.~Sun, I.~Kadota, R.~Talak, and E.~Modiano, ``Age of information: A new metric
  for information freshness,'' {\em Synthesis Lectures on Communication
  Networks}, vol.~12, no.~2, pp.~1--224, 2019.

\bibitem{kam2013age}
C.~Kam, S.~Kompella, and A.~Ephremides, ``Age of information under random
  updates,'' in {\em Proc. IEEE ISIT}, 2013.

\bibitem{huang2015optimizing}
L.~Huang and E.~Modiano, ``Optimizing age-of-information in a multi-class
  queueing system,'' in {\em Proc. IEEE ISIT}, 2015.

\bibitem{inoue2018general}
Y.~Inoue, H.~Masuyama, T.~Takine, and T.~Tanaka, ``A general formula for the
  stationary distribution of the age of information and its application to
  single-server queues,'' {\em IEEE Trans. Inf. Theory}, vol.~65, no.~12,
  p.~8305–8324, 2019.

\bibitem{yin17_tit_update_or_wait}
Y.~Sun, E.~Uysal-Biyikoglu, R.~D. Yates, C.~E. Koksal, and N.~B. Shroff,
  ``Update or wait: How to keep your data fresh,'' {\em IEEE Trans. Inf.
  Theory}, vol.~63, no.~11, pp.~7492--7508, 2017.

\bibitem{bedewy2019minimizing}
A.~M. Bedewy, Y.~Sun, and N.~B. Shroff, ``Minimizing the age of information
  through queues,'' {\em IEEE Trans. Inf. Theory}, vol.~65, no.~8,
  pp.~5215--5232, 2019.

\bibitem{kadota2018scheduling2}
I.~Kadota, A.~Sinha, and E.~Modiano, ``Scheduling algorithms for optimizing age
  of information in wireless networks with throughput constraints,'' {\em
  IEEE/ACM Trans. Netw.}, vol.~27, no.~4, pp.~1359--1372, 2019.

\bibitem{talak2018optimizing}
R.~Talak, S.~Karaman, and E.~Modiano, ``Optimizing information freshness in
  wireless networks under general interference constraints,'' in {\em Proc. ACM
  MobiHoc}, 2018.

\bibitem{maatouk2020optimality}
A.~Maatouk, S.~Kriouile, M.~Assaad, and A.~Ephremides, ``On the optimality of
  the whittle's index policy for minimizing the age of information,'' {\em
  arXiv preprint arXiv:2001.03096}, 2020.

\bibitem{tripathi2019whittle}
V.~Tripathi and E.~Modiano, ``A whittle index approach to minimizing functions
  of age of information,'' in {\em Proc. IEEE Allerton}, 2019.

\bibitem{tripathi2021online}
V.~Tripathi and E.~Modiano, ``An online learning approach to optimizing
  time-varying costs of aoi,'' in {\em Proc. ACM MobiHoc}, 2021.

\bibitem{jhun2018age}
P.~R. Jhunjhunwala and S.~Moharir, ``Age-of-information aware scheduling,'' in
  {\em Proc. IEEE SPCOM}, 2018.

\bibitem{farazi2018age}
S.~Farazi, A.~G. Klein, J.~A. McNeill, and D.~R. Brown, ``On the age of
  information in multi-source multi-hop wireless status update networks,'' in
  {\em Proc. IEEE SPAWC}, 2018.

\bibitem{AoI_measure_1}
C.~Sönmez, S.~Baghaee, A.~Ergişi, and E.~Uysal-Biyikoglu,
  ``Age-of-information in practice: Status age measured over {TCP/IP}
  connections through {WiFi}, ethernet and {LTE},'' in {\em Proc. IEEE
  BlackSeaCom}, 2018.

\bibitem{AoI_measure_3}
H.~B. Beytur, S.~Baghaee, and E.~Uysal, ``Measuring age of information on
  real-life connections,'' in {\em Proc. IEEE SIU}, 2019.

\bibitem{shreedhar2018acp}
T.~Shreedhar, S.~K. Kaul, and R.~D. Yates, ``{ACP}: Age control protocol for
  minimizing age of information over the internet,'' in {\em Proc. ACM
  MobiHoc}, 2018.

\bibitem{AoI_Wierman}
S.-H. Tseng, S.~Han, and A.~Wierman, ``Trading throughput for freshness:
  Freshness-aware traffic engineering and in-network freshness control,'' {\em
  arXiv preprint arXiv:2106.02156}, 2021.

\bibitem{AoI_SDR}
Z.~Han, J.~Liang, Y.~Gu, and H.~Chen, ``Software-defined radio implementation
  of age-of-information-oriented random access,'' in {\em Proc. IEEE IECON},
  2020.

\bibitem{ayan2021experimental}
O.~Ayan, H.~Y. {\"O}zkan, and W.~Kellerer, ``An experimental framework for age
  of information and networked control via software-defined radios,'' in {\em
  Proc. IEEE ICC}, 2021.

\bibitem{kadota2021age}
I.~Kadota and E.~Modiano, ``Age of information in random access networks with
  stochastic arrivals,'' in {\em Proc. IEEE INFOCOM}, 2021.

\bibitem{kadota2021wifresh}
I.~Kadota, M.~S. Rahman, and E.~Modiano, ``{WiFresh}: Age-of-information from
  theory to implementation,'' in {\em Proc. IEEE ICCCN}, 2021.

\bibitem{AoI_management}
M.~Costa, M.~Codreanu, and A.~Ephremides, ``On the age of information in status
  update systems with packet management,'' {\em IEEE Trans. Inf. Theory},
  vol.~62, no.~4, pp.~1897--1910, 2016.

\bibitem{NTP}
D.~Mills, J.~Martin, J.~Burbank, and W.~Kasch, ``Network time protocol version
  4: Protocol and algorithms specification,'' RFC 5905, 2010.

\bibitem{tal2020accurate}
E.~Tal and S.~Karaman, ``Accurate tracking of aggressive quadrotor trajectories
  using incremental nonlinear dynamic inversion and differential flatness,''
  {\em IEEE Trans. Control Syst. Technol.}, vol.~29, no.~3, pp.~1203--1218,
  2021.

\end{thebibliography}

\end{document}